\documentclass[aps,pre,reprint,superscriptaddress,amsmath,showpacs]{revtex4-1}

\usepackage{graphicx}
\usepackage{xcolor}
\usepackage{placeins}
\usepackage{hyperref}

\usepackage{longtable}
\begin{document}

\title{Inequality in economic shock exposures across the global firm-level supply network} 
	
\author{Abhijit Chakraborty}
\thanks{Equal contribution}
\affiliation{Complexity Science Hub Vienna, A-1080 Vienna, Austria}

\author{Tobias Reisch}
\thanks{Equal contribution}
\affiliation{Section for Science of Complex Systems, CeMSIIS, Medical University of Vienna, A-1090 Vienna, Austria}
\affiliation{Complexity Science Hub Vienna, A-1080 Vienna, Austria}

\author{Christian Diem}
\affiliation{Complexity Science Hub Vienna, A-1080 Vienna, Austria}
\affiliation{Institute for Finance, Banking and Insurance, Vienna University of Economics and Business, A-1020 Vienna, Austria}

\author{Stefan Thurner}
\email{Corresponding author: stefan.thurner@meduniwien.ac.at}
\affiliation{Section for Science of Complex Systems, CeMSIIS, Medical University of Vienna, A-1090 Vienna, Austria} 
\affiliation{Complexity Science Hub Vienna, A-1080 Vienna, Austria}
\affiliation{Santa Fe Institute, Santa Fe, NM 85701, USA}



\begin{abstract}  
		For centuries, national economies created wealth by engaging in international trade and production.
		The resulting international supply networks not only increase wealth for countries, but also create systemic risk: 
		economic shocks, triggered by company failures in one country, may propagate to other countries.  
		Using global supply network data on the firm-level, we present a method to estimate a country's exposure to direct and indirect economic losses caused by the failure of a company in another country. 
		We show the network of systemic risk-flows across the world. 
		We find that rich countries expose poor countries much more to systemic risk than the other way round.
		We demonstrate that higher systemic risk levels are not compensated with a risk premium in  GDP, 
		nor do they correlate with economic growth. 
		Systemic risk around the globe appears to be distributed more unequally than wealth.
		These findings put the often praised benefits for developing countries from globalized production
		in a new light, since they relate them to the involved risks in the production processes. 
		Exposure risks present a new dimension of global inequality, that most affects the poor in supply shock crises. It becomes fully quantifiable with the proposed method. 
		\end{abstract}

	\flushbottom
	\maketitle
	\thispagestyle{empty}

	Interconnected supply chains forming complex networks span the globe as a consequence of centuries of globalization \cite{chase2000trade}. International production and trade played an essential role in increasing economic growth~\cite{frankel1999does,pavcnik2002trade,ventura2005global}, reducing global income inequality, especially due to above average growth in China and South Asia~\cite{firebaugh2004accounting}, and has been argued to positively affect sustainable development~\cite{xu2020impacts}.
	Previous studies also argued that international production 
	can have negative effects. Globalization allowed firms to execute strongly polluting tasks abroad~\cite{oita2016substantial,zhang2017transboundary,peters2011growth,wiedmann2018environmental} and to outsource labor intensive tasks to countries with weaker labor-rights~\cite{fernandez2013outsourcing,blackstone2021risk} or
	unsafe or violent working conditions~\cite{hobson2013health,bolle2014bangladesh,butt2019supply}.
	International trade creates demand for and facilitates the spread of problematic goods such as conflict minerals~\cite{mizuno2016structure}.
	Importantly, international trade relations also act as direct and indirect transmission channels for economic shocks, such as supply or demand reductions~\cite{gephart2016vulnerability,starnini2019interconnected,klimek2019quantifying,del2020multiplex}. 
	
	Several recent works show that production networks act as transmission channels for economic shocks. A study on natural disasters in the United States~\cite{barrot2016input} finds substantial evidence that shocks propagate from suppliers to customers. Similarly, shock spreading on the firm-level production network subsequent to the Great East Japan Earthquake 2011 has been estimated to have caused a drop in 2.4\% of GDP, more than 100 times more than the immediate direct losses \cite{inoue2019firm}. In a similar study, the effect of the Great East Japan Earthquake has been estimated to have caused a significant 0.47 percentage point decline in Japan's real GDP growth \cite{carvalho2021supply}. In the short term, some firms were affected orders of magnitude stronger. Further, the Great East Japan Earthquake also provided evidence for the cross-country spread of economic shocks. Using firm-level data \cite{boehm2019input} shows that close affiliates of Japanese corporations in the USA experienced large drops in output in the months following the earthquake.
	
	So far, risks of international economic shock propagation are typically studied on highly aggregated flows of goods between countries~\cite{lee2011impact,gephart2016vulnerability,starnini2019interconnected,klimek2019quantifying,del2020multiplex}. 
	However, the intricate topology of the firm-level global supply network has a potentially crucial role for how economic shocks spread across countries. The importance of knowing the detailed network topology for understanding the spreading of shocks has been shown extensively in the finance literature. Shock propagation mechanisms were first developed for financial networks, consisting of banks and liabilities between them \cite{boss2004network,iori2008network, battiston2012debtrank, thurner2013debtrank, diem2020minimal, pichler2021systemic}.
	The potential of a local disruption, e.g. the default of a single bank, causing a system-wide large disruption in a financial network (financial crisis) is called  its \emph{systemic risk}. 
	A full assessment of the exposures created by a financial agent and thus its systemic risk contribution to the system is not just given by its size (i.e. the sum of its direct exposures), but depends crucially on its position in the network. 
	Reasonable quantification of systemic risk involves the detailed knowledge of financial networks; A particularly practical quantity (network centrality measure) is the DebtRank \cite{battiston2012debtrank,thurner2013debtrank,bardoscia2015debtrank,poledna2015multi} that relates the failure of a bank to the caused systemic losses. 
	
	\begin{figure*}[ht]
		\centering
		\includegraphics[width=0.99\linewidth]{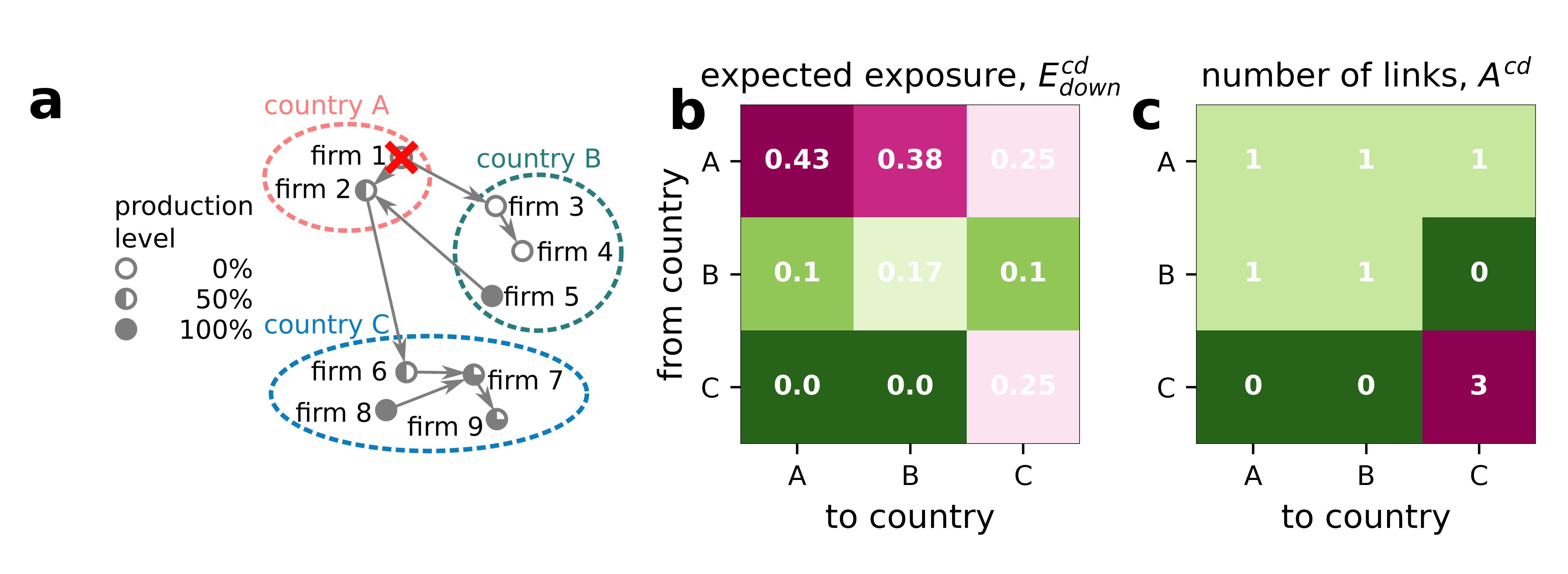}
		\caption{Shock propagation on a firm-level global supply network. 
			(a) A cascade of production reduction in a toy economy of three nations and nine firms, subsequent to the default of firm 1, marked by a red cross. The filling of the nodes (pie chart) indicates their remaining production level after the shock propagated. The production of a firm is reduced proportional to the relative amount of inputs not available.
			(b) Exposures to production losses between the three countries from single firm failures. The expected downstream exposure country $c$ poses to country $d$, $E^{cd}_{down}$,  is measured as the expected production loss in country $d$ if a random firm in country $c$ defaults. Green (magenta) indicates low (high) exposure to expected production losses.
			(c) Number of links between two countries, $A^{cd}$. Green (magenta) indicates low (high) number of direct links.
			The country-adjacency matrix $A^{cd}$ is a poor predictor of $E^{cd}_{down}$, highlighting the relevance of the of intra-country network topology.}
		\label{fig:toyexample}
	\end{figure*}	
	
	Only recently, these ideas were applied and generalized to the real economy and supply networks 
	~\cite{fujiwara2016debtrank}. In networks formed by the supply-demand interactions between firms, the default of a single company may cause --through cascading dynamics-- disruptions in large parts of the system. A corresponding Economic Systemic Risk Index was developed specifically for production networks~\cite{diem2021quantifying}. In a similar direction, agent-based-modelling approaches for estimating the economic cost of the failure of a group of firms, known as regional adaptive input output models, have been developed in the context of (natural) disaster impact assessment~\cite{hallegatte2008adaptive,inoue2019firm,krichene2020model,markhvida2020quantification}.

	Works of this kind demonstrate that network effects on the firm level are relevant and are too large to be ignored. The standard approach of quantifying exposures is the input-output (IO) analysis on the sector-level. There, all networks below the  sector level are ignored.
	Illustrative examples of shock spreading on the single firm level are the global shortage in hard drives subsequent to the 2011 flood in Thailand~\cite{haraguchi2015flood}, or the ongoing shortage in computer chips~\cite{sweney2021global,isidore2021car}.
	In the wake of the COVID-19 crisis, the distress of one firm --\emph{Taiwan Semiconductor Manufacturing Company, Limited}-- lead to production interruptions and layoffs in companies on the other side of the globe, e.g. in European and US car manufacturers \cite{waldersee2021chip,williams2021car}.

	The underlying principle of the above approaches is the understanding of shock propagation on the 
	underlying economic networks in combination with the corresponding actual economic mechanics.
	Note that there are important differences between financial and production networks. 
	In the former links (assets and liabilities) are stock quantities, while in the latter links represent traded goods and services which are 
	flow quantities. In this work we focus on production networks and their systemic risks only. 
	In Fig.~\ref{fig:toyexample}a we show the situation for a cascading failure in an international production network subsequent to the failure of an initial node, firm 1, marked by the red cross. Because firm 1 will no longer produce any goods, it cannot supply inputs to firms 2 and 3, so they have to reduce their production as well and cannot supply to their customers (ignoring potential substitution effects). Iterating this logic leads to a new stable configuration of reduced production levels shown in Fig.~\ref{fig:toyexample}a. The filling of the nodes mark the reduction in economic activity (production). In the case of a supply shock, we call this mechanism the \emph{downstream} propagation of shocks or the \emph{downstream cascade}. The same logic applies to demand reductions that propagate \emph{upstream}, or equivalently, cause an \emph{upstream cascade} (not shown).
	
	\begin{figure*}[ht]
		\centering
		\includegraphics[width=0.85\textwidth]{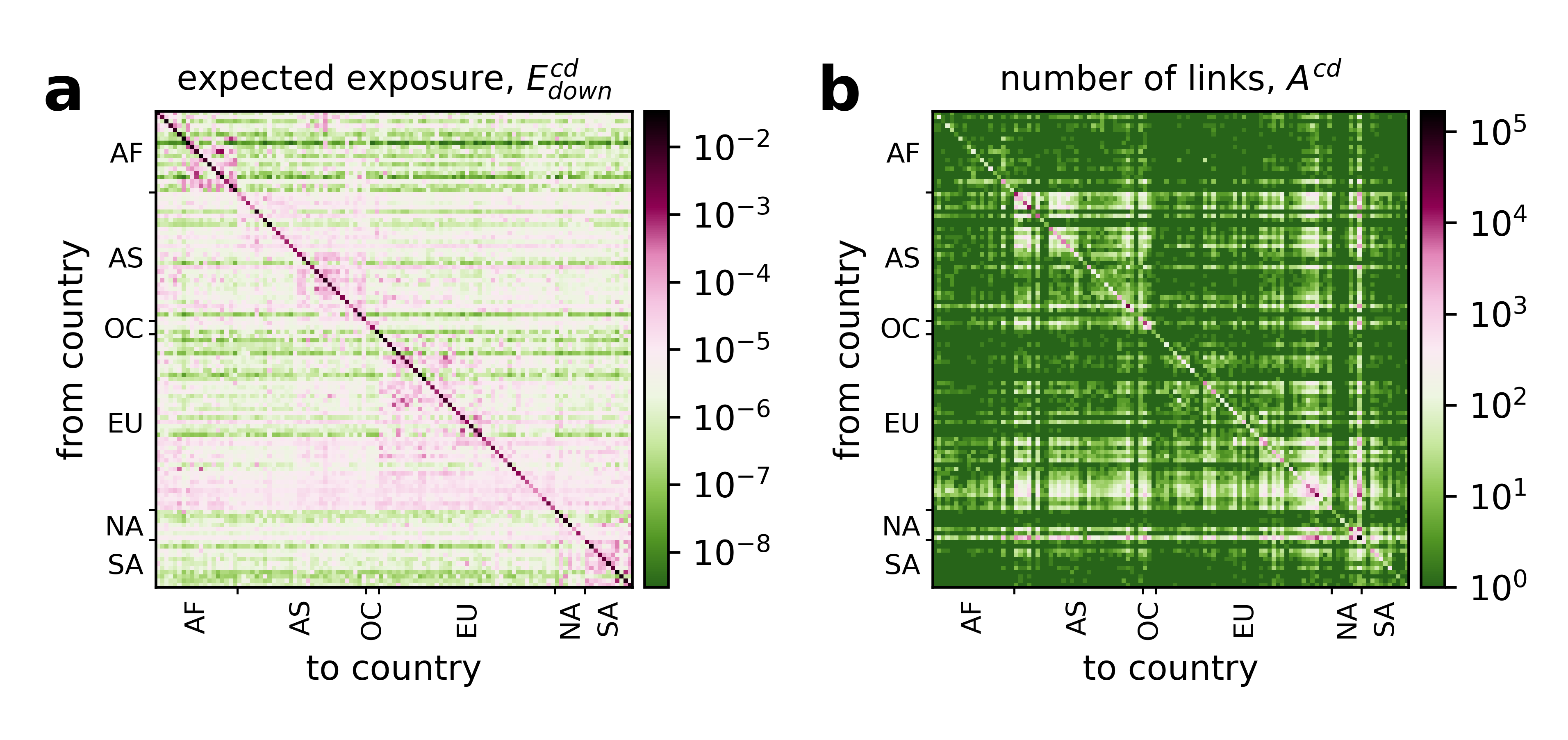}
		\caption{Exposures between countries in the international supply network, aggregated to the country level. 
			(a) Expected fraction of the economy affected in country $d$, following a (random) firm default in country $c$, $E^{cd}_{down}$ (in logarithmic scale). Lines and columns are sorted by continent and region, Africa (AF), Asia (AS), Oceania and Australia (OC), Europe (EU), North America (NA), and South America (SA). Exposures are highest along the diagonal. One can clearly see a regional block structure indicating large exposures within regions and economic blocks. This is especially visible for AS, SA, AF. We find prominent horizontal  lines (magenta), indicating that some countries create exposure well beyond their geographic region, exposing almost all other countries.
			A lager representation with all values and individual country numbering can be found in SI Fig. \ref{fig:heatmap_large}.  
			(b) Country-level adjacency matrix, $A^{cd}$, showing the number of firm-firm inks from country $c$ to country $d$, in the same order as in panel (a). To avoid problems with the logarithmic scale, we add 1 to every entry before taking the log.}
		\label{fig:heatmap_2}
	\end{figure*}
	
	The argument that all involved parties benefit from international trade dates back to David Ricardo's theory of \emph{comparative advantage}~\cite{ricardo1891principles}.
	However, because trade links can act as channels for shock transmission in production networks, it exposes firms and countries to risks they usually can neither assess, nor control. Therefore, quantitative measures are needed that allow us to compare the benefits with the inherent production risks of a globalized economy. 
	Here we develop a novel measure to quantify a country's exposure to economic shocks from international production and trade, based on a microscopic shock propagation mechanism. It is an estimate for the expected economic loss a country is exposed to if an arbitrary firm fails in another country. 
	As such it is also a novel measure for a country's resilience to supply network shocks originating in other countries. For this, we define the direct exposure to be the production losses caused by a (temporary) production failure of a direct supplier (customer) firm. Indirect
	--or, equivalently, higher order-- 
	exposures result from the propagation of the direct shocks along indirect supply relations (supplier of a supplier and so forth). In the following, the term `exposure' refers to the sum of direct and indirect exposures. 
	We can then quantify differences in the countries' risk exposures, and discover how these exposures are distributed across the globe. The estimated exposures can then be put into perspective with measures of gains from globalized production and trade.
	We focus on an accurate description of international economic shock propagation and compare the exposure to direct and indirect production losses with GDP growth, as a straight forward proxy for profits from globalized production and trade.

	We use a global firm data set, see Materials and Methods, as a starting point to reconstruct  an international firm-firm supply network. We use it to compute a number different exposures, based on the concept of DebtRank, see SI Text 1. While DebtRank $R_i$ in its basic form quantifies the systemic relevance of a node by aggregating the losses it could cause in the entire network, here we generalize it such that the systemic relevance of firms to specific countries can be quantified. We thus adapt the cascading mechanism underlying $R_i$ on supply networks in~\cite{fujiwara2016debtrank} to quantify the effect of the failure of each single firm  on every national economy. We define the following quantities:

	The \emph{Country-Firm Exposure}, $E_i^d$, of country $d$ to firm $i$, quantifies the fraction of the national production of country, $d$,  that is lost in case of firm $i$'s failure. This includes direct and indirect shocks transmitted through the network; for details, see Materials and Methods. 
	We define the \emph{Country-Country Exposure}, $E^{cd}$, as the expected exposure of a country $d$ to production losses originating from random firm failures in country $c$.  Assuming that the default probability for firm $i$ is $p_i$, it is calculated by forming the expectation values over all production losses,  $E^{cd} = \sum_{i \in \mathcal{C}^c} p_i E_i^d / |\mathcal{C}^c |$, where $i$ is located in country $c$ and $|\mathcal{C}^c |$ is the number of firms in country $c$. Lacking information on firm default probabilities, we assume equal $p_i$ for all firms in a country. 
	Note that the expected exposure, $E^{cd}$, measures the robustness of country $d$ to shocks from country $c$, low (high) values imply high (low) robustness.
	$E^{cd}$ measures a country's relative production reduction. For absolute numbers, we define a country's \emph{Country-Country Exposed Value}, $V^{cd}$, by multiplying the relative expected exposure with the economic size $q^c$ of the affected country, $V^{cd} = k^d E^{cd}$.  We approximate a firm's size by its number of links, $k_i$, and the economic size of a country by the sum of its firm-sizes, the total degree, $k^c = \sum_{i\in \mathcal{C}^c} k_i$.
	To  distinguish between up- and downstream effects, we define the above measures for both up- ($E_{up}^{cd}$) and downstream  ($E_{down}^{cd}$) separately.
	We define the total (imported) exposure as $E^d = \sum_c E^{cd}$. 
	
	We exemplify the quantities by calculating them for the toy example in Fig.~\ref{fig:toyexample}a. The default of firm 1 causes production in country B to drop by $E_1^B = 75\%$. The default of firm 2, as shown in SI Fig. \ref{fig:SI_toyexample_firm2}, SI Text 2, has no effect on country B, $E_2^B = 0\%$. If we randomly pick a firm in country A, now the expected country exposure is the average of the two firms $E^{AB} = (0+0.75)/2 = 0.375$.  Figure~\ref{fig:toyexample}b shows the $E^{cd}_{down}$ matrix for the toy network in panel a; dark magenta means high, green means low exposures, respectively. Country A creates the most risk, affecting B and C;  C creates the lowest, affecting only itself. For every row the highest value is found in the diagonal, highlighting that firms within countries expose each other stronger among themselves than between countries. The comparison of $E^{cd}_{down}$ in Fig.~\ref{fig:toyexample}b with the number of links between countries $A^{cd} = \sum_{i\in\mathcal{C}^c,j\in\mathcal{C}^d} A_{ij}$ in Fig.~\ref{fig:toyexample}c indicates that the systemic risk spreading dynamics contains a lot of effects that are not visible when only considering the links between countries. Both for directly and indirectly connected countries, much of the heterogeneity in shock spreading is lost when ignoring the firm-level network topology. The information loss by aggregation is best seen when comparing different realizations of the firm-firm level supply network, $A_{ij}$, that have the same country-level aggregation, $A^{cd}$. For example, if we rewire the outgoing link of firm 5, such that it now supplies firm 2 instead of firm 1, we preserve the number of links between countries $A^{cd}$, but the exposure of countries A and C to B would be substantially larger than compared to the situation shown in Fig.~\ref{fig:toyexample}.

	\section*{Results}

	First we analyze the network structure of the expected exposure between countries, $E^{cd}_{down}$.
	We find that \emph{country country exposures} cluster in geographic regions.
	We see this by sorting the countries according to regions and continents in Fig.~\ref{fig:heatmap_2}.  Figure~\ref{fig:heatmap_2}a shows  $E^{cd}_{down}$; panel b shows the number of firm-firm links, $A^{cd}$, between  countries $c$ and $d$. Quantities are in logarithmic scale.
	The strongest connectivity and exposure is found within countries, as seen in the high values along the diagonal in the $A^{cd}$ and $E^{cd}_{down}$  matrices. The effect is much stronger for $E^{cd}_{down}$ than for $A^{cd}$.
	
	Economic exposure can be expected to be strong between countries in the same geographic or economic region, such as the European Union or the United States-Mexico-Canada Agreement. 
	Figure~\ref{fig:heatmap_2}a $E^{cd}_{down}$ clearly shows a corresponding block-diagonal structure, highlighting closely connected regions on all continents, even differentiating between sub-regions, such as northern and sub-Sahara Africa. 
	Further prominent blocks are seen in the middle east and south Asia, eastern Europe, in contrast to northern, southern and western Europe which expose almost all countries, and South America.
     The magenta colored horizontal lines in Fig.~\ref{fig:heatmap_2}a indicate that some countries export downstream exposures to almost all other countries. They are particularly notable for a block of western European, several Asian and North American countries. Rich economies appear to expose countries beyond their own region.
	For a more detailed visualization and discussion  of $E^{cd}_{down}$, containing single country indices,  see SI Fig.~\ref{fig:heatmap_large} and SI Text 3.
	For $A^{cd}$, presented in Fig.~\ref{fig:heatmap_2}b, these structures are only barely visible, except faintly for parts of Europe (EU), Asia (AS) or South America (SA).  In SI Text 4, we investigate this aspect further by comparing the exposed value $V^{cd}$ to the average number of links from firms in country $c$ to country $d$, $\bar{k}^{cd}$. Although $V^{cd}$ and $\bar{k}^{cd}$ are highly correlated (Pearson's $r = 0.93$, $p<10^{-15}$), we find variations of up to two orders of magnitude in $V^{cd}$ for a given $\bar{k}^{cd}$, see SI Fig.~\ref{fig:d_vs_DRcd}.
	
	\begin{figure}[htbp]
		\centering
		\includegraphics[width=\linewidth]{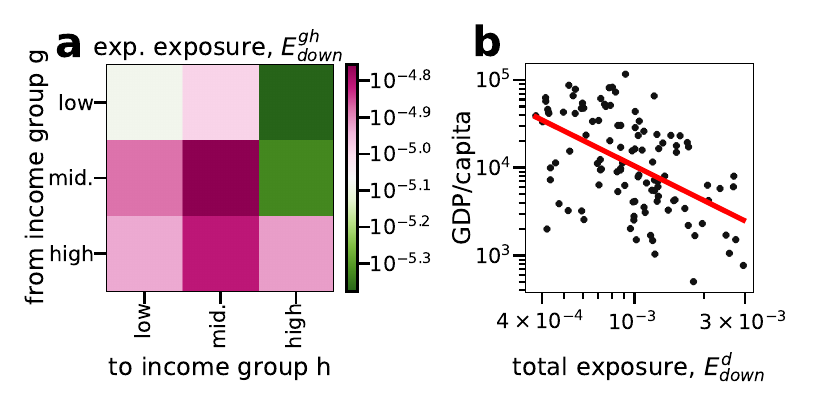}
		\vspace*{-8mm}
		\caption{Exposures between high-, middle, and low income coutries. 
			(a) Exposure ($\log$-scale) between firms separated into low-, middle- and high-income groups based on their country's GDP per capita. Firms in middle-income countries experience the largest amount of exposure and high-income countries are mostly exposed to other high-income countries. Firms in low income countries are much more exposed than they expose others. SI Table \ref{tab:SI:country_names} lists the income group for each country.
			(b) GDP per capita plotted against total exposure, $E^d_{down}$. A significant negative correlation of $r=0.52, p<10^{-8}$ highlights that higher exposure is connected to lower income per capita. The red line represents the log-log ordinary least squares regression fit to the data of form $y \sim x^{-1.31}$.
		}
		\label{fig:GDPassortativity}
	\end{figure}
	
	The obvious asymmetry in $E^{cd}_{down}$  in Fig.~\ref{fig:heatmap_2}a suggests large differences in how systemic risk is distributed around the globe and that exposure to economic losses is usually not reciprocal. 
	In Fig.~\ref{fig:GDPassortativity}a we change the aggregation scale to three income groups, containing the same number of countries, based on their GDP per capita --low, middle and high. We plot the respective $E^{gh}$, where $g$ and $h$ denote the respective country income groups. Three facts become obvious.  
	First, firms in high income countries export much more distress to middle and lower income groups than they import from these. 
	Second, firms in high and middle income countries are most exposed to firms in countries from their own respective groups. Firms in low-income countries receive relatively little risk from their own income group but more from middle and high income countries, indicating that these countries' economies are more exposed to the wealthier trading partners. 
	Third, the  dark color of the column for middle income countries in Fig.~\ref{fig:GDPassortativity}a indicates that they are exposed to most of the risk.
	
	In Fig~\ref{fig:GDPassortativity}b we plot the total exposure, $E^d_{down}$, versus GDP per capita showing an anti-correlation (Pearson $r=-0.52$, $p<10^{-8}$). This indicates that countries with a low GDP per capita are more exposed than  countries with a higher. To control for  confounding factors, we perform a multivariate linear regression, where GDP per capita is the dependent variable and $E^{cd}_{down}$, GDP, imports, exports, imports per capita and exports per capita are the independent variables. The results of the regression analysis are shown SI Tab. \ref{table:1} and SI Text 5 and indicate that only $E^{cd}_{down}$ and exports per capita have a statistically significant influence on the dependent variable. The model explains 67\% of the variance and is highly significant (adjusted $R^2=0.67$, $p<10^{-15}$). 
	
	Generally, high risks are thought to be compensated with higher expected returns~\cite{sharpe1964capital, sharpe1966mutual, fama2015five}. We check for the existence of such a ``risk premium'' by comparing $E^{d}$ with the average annual growth of GDP per capita in the respective country over the past twenty years. 
	A high growth rate has been causally connected to successful export strategies \cite{van2015international,ramzan2019impact}.
	SI Figure \ref{fig:SI:growthvsexpsure}, in SI Text 6,  shows that there is little to no correlation between the variables (insignificant Pearson correlation of $r=0.08, p = 0.37 $). Countries seem not to be compensated for taking systemic risk. In SI Text 6 we check for the robustness of this result by comparing with the average annual GDP growth over 5 and 10 years. 
	
	\begin{figure}[htb]
		\centering
		\includegraphics[width=0.65\linewidth]{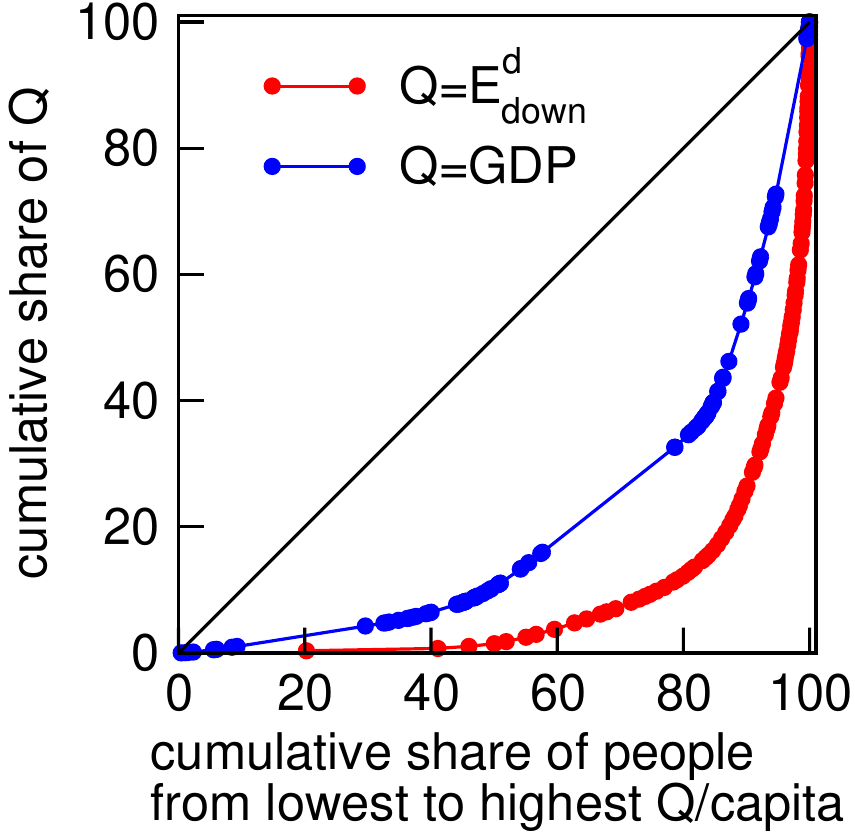}
		\caption{Lorenz curves for total exposure and GDP of all countries. The red (blue) line shows the proportion of risk (GDP) that is associated to the lowest exposed (poorest) x\% of people globally. For perfectly equally distributed exposures (wealth) the curve would coincide with the diagonal. Exposure to economic shock is obviously distributed more unequally than GDP. We find Gini coefficients of 0.83 and 0.59 for $E^d_{down}$ and GDP, respectively. 
			Systemic risk exposures represent a new aspect of inequality.
		}
		\label{fig:inequality}
	\end{figure}	
	
	To find out how equally exposures are  distributed  across the globe, in Fig.~\ref{fig:inequality} we plot the respective Lorenz curve, i.e. the cumulative share of population (sorted by their countries' $E^d_{down}$ per capita) versus their cumulative share of global expected exposure, $E^d_{down}$. 80\% of population that are least exposed to risk are exposed to only around 10\% of all risk, or vice-versa, 20\% of the most exposed population carries around 90\% of the risk. For comparison, in Fig.~\ref{fig:inequality} we also show the Lorenz curve for the GDP. Clearly, GDP is distributed somewhat more equally than the $E^d_{down}$, which is also shown by the Gini coefficient. We find Gini coefficients for the GDP and total exposure $E^d_{down}$ to be $0.59$ and $0.83$, respectively. 
	
	So far we only considered \emph{downstream shock propagation} by simulating supply shortages.
	The analysis for the \emph{upstream} cascade is discussed in SI Text 7. We find similar structures with respect to the expected upstream exposure $E^{cd}_{up}$ matrix (analogously defined as $E^{cd}_{down}$) in SI Fig.~\ref{fig:heatmap_up}. The distribution of upstream exposure by income group in SI Fig.~\ref{fig:GDPassortativity_up} reveals a different picture than for downstream risk. Most risk is between low- and middle-income countries and high-income countries neither create nor are exposed to risk. The respective Gini coefficient is 0.81 and therefore higher than for GDP, again to the disadvantage of poor countries (Pearson's correlation between GDP per capita and $E^{cd}_{up}$ is $r=-0.20$, $p<0.04$), see SI Fig.~\ref{fig:inequality_up}.

	\section*{Discussion}
	We present a method to quantify the exposure of countries to production losses caused by firm defaults in other countries based on global firm-level supply network data. We introduce the \emph{expected Country-Country Exposure}, $E^{cd}$, as the expected relative loss of production in country $d$ after a firm-failure in country $c$. The method enables us to demonstrate that exposures to other countries is highly  structured on a regional level, and that high income countries expose a large fraction of the globe to economic losses. Low income countries are disproportionately strongly affected by high exposure values. Somewhat contrary to intuition, it seems that higher exposure is not positively correlated with higher gains in GDP growth rates in recent decades. Global economic exposure of the type discussed is distributed more unequal than income per capita. 
	
	The presented metric confirms the intuition that expected exposures of countries are highest to firms within their own economies, and on the international level exposures are strongly influenced by geography. When countries are sorted by continent, the country exposure matrix, $E^{cd}_{down}$, shows a prominent block-diagonal structure. This is not unexpected, since it is known that exposure increases with connectivity, see SI Text 4, and because trade intensity decreases with distance~\cite{anderson2011gravity}.
	Several wealthy countries create significant exposures beyond their region and thus export systemic risk to other countries.
	We show that the number of trade links between countries are bad predictors for the direct and indirect exposures between countries. Similarly, company size has been found to be a bad predictor for firms' systemic riskiness in supply networks ~\cite{diem2021quantifying,fujiwara2016debtrank}. 
	
	To understand if the asymmetry of few countries exposing many countries and many countries exposing only a few is related to their average income level, we separate countries into low-, middle-, and high income groups. We find that firms in high income countries have a privileged position, in the sense that they are only exposed to risk from other high income countries, but expose all other groups. On the other hand, low income countries hardly expose others, while they themselves are exposed strongly. Middle income countries are sizeably exposed to all three income groups. The asymmetry of exposure does \emph{a priori} not provide any indication whether rich or poor countries have higher total exposure. On the one hand, poor countries could be more dependent on inputs from rich countries, on the other hand rich countries could depend more on highly integrated global value chains and, hence, be more exposed to systemic risk. A regression analysis solves the puzzle and  we find that GDP per capita correlates negatively ($r=0.52, p<10^{-8}$) and, hence, poor countries have a higher total exposure. The presented results on the spreading of supply shocks, are complemented in SI Text 7 with the analogous analysis for demand shocks. We find qualitatively similar results. Again, there is high inequality to the disadvantage of low income and advantage of high income countries. With the currently available data we find no evidence for a ``risk premium'' in the sense that higher exposures would co-occur with significantly higher economic growth rates. The expected exposure is not significantly correlated with the growth of GDP per capita during the past 20 years before the COVID-19 crises ($r=0.08$, $p=0.37$).
	
	We investigate the inequality of exposures across the globe and find that exposure is highly concentrated with a Gini coefficient of 0.83 (compared to 0.59 for GDP). This additional dimension of global inequality adds to a number of other forms of inequality such as inequality with respect to health, (formal) education, and wealth~\cite{goesling2008three}. Inequality impedes economic growth \cite{barro1999inequality,oecd2015init} and has been associated with being one of the driving factors for the collapse of societies \cite{turchin2016ages}. In their Agenda 2030 the United Nations set to ``reduce inequality within and among countries'' (SDG 10) as one of the 17 sustainable development goals (SDGs) \cite{un2015sdg}. Future research should investigate if the network structure of the global supply network locks in already existing inequalities. For social networks it has been shown that their structure can procreate inequality~\cite{dimaggio2012network,karimi2018homophily}. Similarly, our results suggest that structural inequality in  production and trade networks between countries is significant and should be taken into consideration in future efforts to fight inequality in its various forms. Since exposure inequality arises from the structure of the supply network on the firm-level, it is important to understand the processes that let firms from different income countries enter into production and trade relations and how this could happen with creating less risk exposure to poorer countries. Strategies of incentivizing firms to become systemic risk sensitive could be a starting point \cite{leduc2017incentivizing,poledna2016elimination}.
	
	The presented work shows how to calculate international direct and indirect economic exposures between countries that is methodically based on previous work on financial and supply networks~\cite{battiston2012debtrank,fujiwara2016debtrank}. The present work is only a first  step in the direction of quantifying systemic risk flows around the globe and has several limitations that need improvement. First, our approach disregards completely the nature of the products. An accurate description  shock propagation must include the type of the goods and their firm-level production functions. Only then can a supply network be considered a realistic production network. At the moment, the shock spreading dynamics are practically limited to the special case of linear production functions that are typically underestimating the actual shock spreading risks~\cite{diem2021quantifying}. Second, the current dataset lacks information on the firm's revenues and the traded volumes. In SI Fig.~\ref{fig:weight}, in SI Text 8, we show that GDP, imports, and exports all have a high correlation with a country's total degree, outdegree, and indegree. More detailed supply network information would improve the quality of our results by capturing relative effects in the supply network propagation. Third, the dataset is compiled by firms centered in the US, entailing a potential reporting bias. To ensure the robustness of our results, we perform the same study for states in the USA. Although the heterogeneity in the US is much lower, we find similar trends, such as high inequality to the benefit of rich states, for details see SI Text 9. Fourth, the shown results were derived for the spreading of supply and demand shocks separately. We find qualitatively similar results; there is high inequality to the disadvantage of low income and advantage of high income countries. However, it would be desirable to design a measure that is able to capture up- and downstream risk simultaneously, similar to the ESRI quantity recently presented in~\cite{diem2021quantifying}.  
	Fifth and finally, we assume the same default probability, $p_i$, for every firm and simulate the ensuing cascade of production interruptions. This is certainly unrealistic, and heterogeneous default probabilities of companies should be taken into account, as well as information on correlated shocks, e.g. due to natural disasters or climate change. However, this is presently practically impossible due to unavailability of corresponding data.

	The presented work allows us to derive three immediate policy implications: First, we conclude that the (global) distribution of exposure to economic shocks must be traced on granular representations of the underlying networks. For this a global effort on collecting and monitoring the according data is necessary. This would allow to anticipate and prepare for globally spreading supply shocks. Second, since inequality is deeply embedded in the economic structure of the global supply network, future efforts to reduce inequality, such as the goals formulated in SDG 10 in the United Nations' Agenda 2030, must include the systematic restructuring of the global production network to spread exposure  risky more fairly. Third, since the developed index quantifies the spread of exposure to economic shocks between countries. The employed firm-level resolution allows us to straight forwardly adapt systemic risk management methods for national economies. A possible incentive scheme could be an appropriately adapted systemic risk tax \cite{leduc2017incentivizing,poledna2016elimination} for international supply networks.
	
	\section*{Materials and Methods}
	\subsection*{Data}
	We use supplier-customer data from the web site of Standard \& Poor's (S\&P) Capital IQ platform for the year $2017$. The data is comprised of firm identification (ID), name, location, primary industry, and sector as node information. Using the Global Industry Classification Standard (GICS), Morgan Stanley Capital International and S\&P grouped the firms into 11 sectors and 158 primary industries. Firms are distributed over 206 countries. The data contains information on 1,403,807 business relationships between firms, such as supplier, creditor, franchisor, licensor, landlord, lessor, auditor, transfer agent, investor relations firm, and vendor. Most relations are of type supplier or creditor (69\% supply links, 31\% creditor links. The supplier type implies that a firm provides goods or services to other firms, the creditor type indicates that a firm lends money to other firms. For a detailed description of the dataset we refer to \cite{chakraborty2020testing}. Because we are interested in the flow of goods and services between countries, in the following we restrict our analysis to the 968,627 supplier relations.
	
	We preprocess the data in the following way. We remove all firms that do not have locations or sector classification information. To avoid misleading results from countries with too few firms, we only consider firms from those countries where the total number of firms exceeds $30$. Further, we remove firms from Barbados, Bermuda, British virgin Island, Channel Island, Gibraltar, Monaco, and Cayman Island as these are known for having considerable numbers of offshore firms. We construct an unweighted network with an adjacency matrix, $A_{ij}=1$, if there exists a link between firm $i$ and $j$ and $A_{ij}=0$, otherwise. After removing isolated nodes, parallel links, and self-loops, the network contains $N =230,970 $ firms and $L=660,701$ binary, directed links. 
	
	To explain the variation in the amount of distress propagation for different countries, we investigate the relation between 
	distress and certain economic indicators characterizing an economy. We collect data on per capita gross domestic product, per capita exports and per capita imports for the year 2018 in current U.S. dollars from the website of the World Bank \url{https://data.worldbank.org}. 
	
	\subsection*{Methods}
	DebtRank $R_i$ is a network centrality measure, initially developed as a financial systemic risk index for investment networks~\cite{battiston2012debtrank}, and recently adopted for supply networks \cite{fujiwara2016debtrank}. In a real economic network $R_i$ corresponds to the overall reduction in economic activity subsequent to the default of an initial firm $i$ and a possible cascade of defaults. Here, we adopt the specific underlying cascading mechanism to design a novel country-level indicator. In SI Text 1 we give a brief review of the DebtRank as used in \cite{fujiwara2016debtrank} and here we describe our specific adaptation.
	
	An intermediate result in the calculation of $R_i$ is the matrix $D$, with elements $D_{ij}$, denoting the distress firm $j$ receives if firm $i$ defaults (the loss in economic production firm $j$ experiences if firm $i$ defaults), see SI Text 1. The column $j$ contains the shocks firm $j$ is exposed to. Row $i$ contains the shocks firm $i$ causes to all other firms $j$ when it defaults. 
	We use $D$ to define the effect of a node's default on a set of nodes $\mathcal{C}$, in our study typically corresponding to the firms in a country $c$, $\mathcal{C}^c$. In the following, lower indices denote nodes (firms) and raised indices denote groups of nodes (countries).

	Because the global supply chain network lacks information on link-weights and node-sizes, we assigned each edge equal weight and associate the node-size $q_i$ with it's degree $q_i=k_i$, where  $k_i$ represents the degree of the $i$-th node. The degree $k_i$ was chosen as a size proxy to be consistent and self contained in the network setting of the used dataset. However, one could also use any other size proxy such as value added, turnover or employees. We define the weight of a country $c$ as the sum of its node weights $$ q^c = \sum_{i \in \mathcal{C}^c} q_i.$$ 
	
	The \emph{Firm-Country Exposure} of node i for country c is 
	$$ E^c_i = \frac{\sum_{j \in \mathcal{C}^c} D_{ij} q_j} {q^c}.$$ 
	Please note that the weight of the initial node is included in the denominator if it is also contained in the set $\mathcal{C}^c$. This allows the interpretation of this value as the fraction of the economy affected in country $c$ after the default of firm $c$.

	\bibliography{bibliography}

\begin{thebibliography}{60}%
\makeatletter
\providecommand \@ifxundefined [1]{%
 \@ifx{#1\undefined}
}%
\providecommand \@ifnum [1]{%
 \ifnum #1\expandafter \@firstoftwo
 \else \expandafter \@secondoftwo
 \fi
}%
\providecommand \@ifx [1]{%
 \ifx #1\expandafter \@firstoftwo
 \else \expandafter \@secondoftwo
 \fi
}%
\providecommand \natexlab [1]{#1}%
\providecommand \enquote  [1]{``#1''}%
\providecommand \bibnamefont  [1]{#1}%
\providecommand \bibfnamefont [1]{#1}%
\providecommand \citenamefont [1]{#1}%
\providecommand \href@noop [0]{\@secondoftwo}%
\providecommand \href [0]{\begingroup \@sanitize@url \@href}%
\providecommand \@href[1]{\@@startlink{#1}\@@href}%
\providecommand \@@href[1]{\endgroup#1\@@endlink}%
\providecommand \@sanitize@url [0]{\catcode `\\12\catcode `\$12\catcode
  `\&12\catcode `\#12\catcode `\^12\catcode `\_12\catcode `\%12\relax}%
\providecommand \@@startlink[1]{}%
\providecommand \@@endlink[0]{}%
\providecommand \url  [0]{\begingroup\@sanitize@url \@url }%
\providecommand \@url [1]{\endgroup\@href {#1}{\urlprefix }}%
\providecommand \urlprefix  [0]{URL }%
\providecommand \Eprint [0]{\href }%
\providecommand \doibase [0]{http://dx.doi.org/}%
\providecommand \selectlanguage [0]{\@gobble}%
\providecommand \bibinfo  [0]{\@secondoftwo}%
\providecommand \bibfield  [0]{\@secondoftwo}%
\providecommand \translation [1]{[#1]}%
\providecommand \BibitemOpen [0]{}%
\providecommand \bibitemStop [0]{}%
\providecommand \bibitemNoStop [0]{.\EOS\space}%
\providecommand \EOS [0]{\spacefactor3000\relax}%
\providecommand \BibitemShut  [1]{\csname bibitem#1\endcsname}%
\let\auto@bib@innerbib\@empty
\bibitem [{\citenamefont {Chase-Dunn}\ \emph {et~al.}(2000)\citenamefont
  {Chase-Dunn}, \citenamefont {Kawano},\ and\ \citenamefont
  {Brewer}}]{chase2000trade}%
  \BibitemOpen
  \bibfield  {author} {\bibinfo {author} {\bibfnamefont {C.}~\bibnamefont
  {Chase-Dunn}}, \bibinfo {author} {\bibfnamefont {Y.}~\bibnamefont {Kawano}},
  \ and\ \bibinfo {author} {\bibfnamefont {B.~D.}\ \bibnamefont {Brewer}},\
  }\href@noop {} {\bibfield  {journal} {\bibinfo  {journal} {American
  Sociological Review}\ }\textbf {\bibinfo {volume} {65}},\ \bibinfo {pages}
  {77} (\bibinfo {year} {2000})}\BibitemShut {NoStop}%
\bibitem [{\citenamefont {Frankel}\ and\ \citenamefont
  {Romer}(1999)}]{frankel1999does}%
  \BibitemOpen
  \bibfield  {author} {\bibinfo {author} {\bibfnamefont {J.~A.}\ \bibnamefont
  {Frankel}}\ and\ \bibinfo {author} {\bibfnamefont {D.~H.}\ \bibnamefont
  {Romer}},\ }\href@noop {} {\bibfield  {journal} {\bibinfo  {journal}
  {American Economic Review}\ }\textbf {\bibinfo {volume} {89}},\ \bibinfo
  {pages} {379} (\bibinfo {year} {1999})}\BibitemShut {NoStop}%
\bibitem [{\citenamefont {Pavcnik}(2002)}]{pavcnik2002trade}%
  \BibitemOpen
  \bibfield  {author} {\bibinfo {author} {\bibfnamefont {N.}~\bibnamefont
  {Pavcnik}},\ }\href@noop {} {\bibfield  {journal} {\bibinfo  {journal} {The
  Review of Economic Studies}\ }\textbf {\bibinfo {volume} {69}},\ \bibinfo
  {pages} {245} (\bibinfo {year} {2002})}\BibitemShut {NoStop}%
\bibitem [{\citenamefont {Ventura}(2005)}]{ventura2005global}%
  \BibitemOpen
  \bibfield  {author} {\bibinfo {author} {\bibfnamefont {J.}~\bibnamefont
  {Ventura}},\ }in\ \href@noop {} {\emph {\bibinfo {booktitle} {Handbook of
  Economic Growth}}},\ Vol.~\bibinfo {volume} {1}\ (\bibinfo  {publisher}
  {Elsevier},\ \bibinfo {year} {2005})\ pp.\ \bibinfo {pages}
  {1419--1497}\BibitemShut {NoStop}%
\bibitem [{\citenamefont {Firebaugh}\ and\ \citenamefont
  {Goesling}(2004)}]{firebaugh2004accounting}%
  \BibitemOpen
  \bibfield  {author} {\bibinfo {author} {\bibfnamefont {G.}~\bibnamefont
  {Firebaugh}}\ and\ \bibinfo {author} {\bibfnamefont {B.}~\bibnamefont
  {Goesling}},\ }\href@noop {} {\bibfield  {journal} {\bibinfo  {journal}
  {American Journal of Sociology}\ }\textbf {\bibinfo {volume} {110}},\
  \bibinfo {pages} {283} (\bibinfo {year} {2004})}\BibitemShut {NoStop}%
\bibitem [{\citenamefont {Xu}\ \emph {et~al.}(2020)\citenamefont {Xu},
  \citenamefont {Li}, \citenamefont {Chau}, \citenamefont {Dietz},
  \citenamefont {Li}, \citenamefont {Wan}, \citenamefont {Zhang}, \citenamefont
  {Zhang}, \citenamefont {Li}, \citenamefont {Chung},\ and\ \citenamefont
  {Liu}}]{xu2020impacts}%
  \BibitemOpen
  \bibfield  {author} {\bibinfo {author} {\bibfnamefont {Z.}~\bibnamefont
  {Xu}}, \bibinfo {author} {\bibfnamefont {Y.}~\bibnamefont {Li}}, \bibinfo
  {author} {\bibfnamefont {S.~N.}\ \bibnamefont {Chau}}, \bibinfo {author}
  {\bibfnamefont {T.}~\bibnamefont {Dietz}}, \bibinfo {author} {\bibfnamefont
  {C.}~\bibnamefont {Li}}, \bibinfo {author} {\bibfnamefont {L.}~\bibnamefont
  {Wan}}, \bibinfo {author} {\bibfnamefont {J.}~\bibnamefont {Zhang}}, \bibinfo
  {author} {\bibfnamefont {L.}~\bibnamefont {Zhang}}, \bibinfo {author}
  {\bibfnamefont {Y.}~\bibnamefont {Li}}, \bibinfo {author} {\bibfnamefont
  {M.~G.}\ \bibnamefont {Chung}}, \ and\ \bibinfo {author} {\bibfnamefont
  {J.}~\bibnamefont {Liu}},\ }\href {\doibase 10.1038/s41893-020-0572-z}
  {\bibfield  {journal} {\bibinfo  {journal} {Nature Sustainability}\ }\textbf
  {\bibinfo {volume} {3}},\ \bibinfo {pages} {964} (\bibinfo {year}
  {2020})}\BibitemShut {NoStop}%
\bibitem [{\citenamefont {Oita}\ \emph {et~al.}(2016)\citenamefont {Oita},
  \citenamefont {Malik}, \citenamefont {Kanemoto}, \citenamefont {Geschke},
  \citenamefont {Nishijima},\ and\ \citenamefont
  {Lenzen}}]{oita2016substantial}%
  \BibitemOpen
  \bibfield  {author} {\bibinfo {author} {\bibfnamefont {A.}~\bibnamefont
  {Oita}}, \bibinfo {author} {\bibfnamefont {A.}~\bibnamefont {Malik}},
  \bibinfo {author} {\bibfnamefont {K.}~\bibnamefont {Kanemoto}}, \bibinfo
  {author} {\bibfnamefont {A.}~\bibnamefont {Geschke}}, \bibinfo {author}
  {\bibfnamefont {S.}~\bibnamefont {Nishijima}}, \ and\ \bibinfo {author}
  {\bibfnamefont {M.}~\bibnamefont {Lenzen}},\ }\href@noop {} {\bibfield
  {journal} {\bibinfo  {journal} {Nature Geoscience}\ }\textbf {\bibinfo
  {volume} {9}},\ \bibinfo {pages} {111} (\bibinfo {year} {2016})}\BibitemShut
  {NoStop}%
\bibitem [{\citenamefont {Zhang}\ \emph {et~al.}(2017)\citenamefont {Zhang},
  \citenamefont {Jiang}, \citenamefont {Tong}, \citenamefont {Davis},
  \citenamefont {Zhao}, \citenamefont {Geng}, \citenamefont {Feng},
  \citenamefont {Zheng}, \citenamefont {Lu}, \citenamefont {Streets} \emph
  {et~al.}}]{zhang2017transboundary}%
  \BibitemOpen
  \bibfield  {author} {\bibinfo {author} {\bibfnamefont {Q.}~\bibnamefont
  {Zhang}}, \bibinfo {author} {\bibfnamefont {X.}~\bibnamefont {Jiang}},
  \bibinfo {author} {\bibfnamefont {D.}~\bibnamefont {Tong}}, \bibinfo {author}
  {\bibfnamefont {S.~J.}\ \bibnamefont {Davis}}, \bibinfo {author}
  {\bibfnamefont {H.}~\bibnamefont {Zhao}}, \bibinfo {author} {\bibfnamefont
  {G.}~\bibnamefont {Geng}}, \bibinfo {author} {\bibfnamefont {T.}~\bibnamefont
  {Feng}}, \bibinfo {author} {\bibfnamefont {B.}~\bibnamefont {Zheng}},
  \bibinfo {author} {\bibfnamefont {Z.}~\bibnamefont {Lu}}, \bibinfo {author}
  {\bibfnamefont {D.~G.}\ \bibnamefont {Streets}},  \emph {et~al.},\
  }\href@noop {} {\bibfield  {journal} {\bibinfo  {journal} {Nature}\ }\textbf
  {\bibinfo {volume} {543}},\ \bibinfo {pages} {705} (\bibinfo {year}
  {2017})}\BibitemShut {NoStop}%
\bibitem [{\citenamefont {Peters}\ \emph {et~al.}(2011)\citenamefont {Peters},
  \citenamefont {Minx}, \citenamefont {Weber},\ and\ \citenamefont
  {Edenhofer}}]{peters2011growth}%
  \BibitemOpen
  \bibfield  {author} {\bibinfo {author} {\bibfnamefont {G.~P.}\ \bibnamefont
  {Peters}}, \bibinfo {author} {\bibfnamefont {J.~C.}\ \bibnamefont {Minx}},
  \bibinfo {author} {\bibfnamefont {C.~L.}\ \bibnamefont {Weber}}, \ and\
  \bibinfo {author} {\bibfnamefont {O.}~\bibnamefont {Edenhofer}},\ }\href@noop
  {} {\bibfield  {journal} {\bibinfo  {journal} {Proceedings of the National
  Academy of Sciences}\ }\textbf {\bibinfo {volume} {108}},\ \bibinfo {pages}
  {8903} (\bibinfo {year} {2011})}\BibitemShut {NoStop}%
\bibitem [{\citenamefont {Wiedmann}\ and\ \citenamefont
  {Lenzen}(2018)}]{wiedmann2018environmental}%
  \BibitemOpen
  \bibfield  {author} {\bibinfo {author} {\bibfnamefont {T.}~\bibnamefont
  {Wiedmann}}\ and\ \bibinfo {author} {\bibfnamefont {M.}~\bibnamefont
  {Lenzen}},\ }\href@noop {} {\bibfield  {journal} {\bibinfo  {journal} {Nature
  Geoscience}\ }\textbf {\bibinfo {volume} {11}},\ \bibinfo {pages} {314}
  (\bibinfo {year} {2018})}\BibitemShut {NoStop}%
\bibitem [{\citenamefont {Fern{\'a}ndez}\ and\ \citenamefont
  {Sotelo~Valencia}(2013)}]{fernandez2013outsourcing}%
  \BibitemOpen
  \bibfield  {author} {\bibinfo {author} {\bibfnamefont {D.~C.}\ \bibnamefont
  {Fern{\'a}ndez}}\ and\ \bibinfo {author} {\bibfnamefont {A.}~\bibnamefont
  {Sotelo~Valencia}},\ }\href@noop {} {\bibfield  {journal} {\bibinfo
  {journal} {Latin American Perspectives}\ }\textbf {\bibinfo {volume} {40}},\
  \bibinfo {pages} {14} (\bibinfo {year} {2013})}\BibitemShut {NoStop}%
\bibitem [{\citenamefont {Blackstone}\ \emph {et~al.}(2021)\citenamefont
  {Blackstone}, \citenamefont {Norris}, \citenamefont {Robbins}, \citenamefont
  {Jackson},\ and\ \citenamefont {Decker~Sparks}}]{blackstone2021risk}%
  \BibitemOpen
  \bibfield  {author} {\bibinfo {author} {\bibfnamefont {N.~T.}\ \bibnamefont
  {Blackstone}}, \bibinfo {author} {\bibfnamefont {C.~B.}\ \bibnamefont
  {Norris}}, \bibinfo {author} {\bibfnamefont {T.}~\bibnamefont {Robbins}},
  \bibinfo {author} {\bibfnamefont {B.}~\bibnamefont {Jackson}}, \ and\
  \bibinfo {author} {\bibfnamefont {J.~L.}\ \bibnamefont {Decker~Sparks}},\
  }\href@noop {} {\bibfield  {journal} {\bibinfo  {journal} {Nature Food}\
  }\textbf {\bibinfo {volume} {2}},\ \bibinfo {pages} {692} (\bibinfo {year}
  {2021})}\BibitemShut {NoStop}%
\bibitem [{\citenamefont {Hobson}(2013)}]{hobson2013health}%
  \BibitemOpen
  \bibfield  {author} {\bibinfo {author} {\bibfnamefont {J.}~\bibnamefont
  {Hobson}},\ }\href {\doibase 10.1093/occmed/kqt079} {\bibfield  {journal}
  {\bibinfo  {journal} {Occupational Medicine}\ }\textbf {\bibinfo {volume}
  {63}},\ \bibinfo {pages} {317} (\bibinfo {year} {2013})},\ \Eprint
  {http://arxiv.org/abs/https://academic.oup.com/occmed/article-pdf/63/5/317/4255408/kqt079.pdf}
  {https://academic.oup.com/occmed/article-pdf/63/5/317/4255408/kqt079.pdf}
  \BibitemShut {NoStop}%
\bibitem [{\citenamefont {Bolle}(2014)}]{bolle2014bangladesh}%
  \BibitemOpen
  \bibfield  {author} {\bibinfo {author} {\bibfnamefont {M.~J.}\ \bibnamefont
  {Bolle}},\ }\href@noop {} {\emph {\bibinfo {title} {Bangladesh apparel
  factory collapse: Background in brief}}},\ \bibinfo {type} {Tech. Rep.}\
  (\bibinfo  {institution} {Congressional Research Service, the Library of
  Congress},\ \bibinfo {year} {2014})\BibitemShut {NoStop}%
\bibitem [{\citenamefont {Butt}\ \emph {et~al.}(2019)\citenamefont {Butt},
  \citenamefont {Lambrick}, \citenamefont {Menton},\ and\ \citenamefont
  {Renwick}}]{butt2019supply}%
  \BibitemOpen
  \bibfield  {author} {\bibinfo {author} {\bibfnamefont {N.}~\bibnamefont
  {Butt}}, \bibinfo {author} {\bibfnamefont {F.}~\bibnamefont {Lambrick}},
  \bibinfo {author} {\bibfnamefont {M.}~\bibnamefont {Menton}}, \ and\ \bibinfo
  {author} {\bibfnamefont {A.}~\bibnamefont {Renwick}},\ }\href@noop {}
  {\bibfield  {journal} {\bibinfo  {journal} {Nature Sustainability}\ }\textbf
  {\bibinfo {volume} {2}},\ \bibinfo {pages} {742} (\bibinfo {year}
  {2019})}\BibitemShut {NoStop}%
\bibitem [{\citenamefont {Mizuno}\ \emph {et~al.}(2016)\citenamefont {Mizuno},
  \citenamefont {Ohnishi},\ and\ \citenamefont
  {Watanabe}}]{mizuno2016structure}%
  \BibitemOpen
  \bibfield  {author} {\bibinfo {author} {\bibfnamefont {T.}~\bibnamefont
  {Mizuno}}, \bibinfo {author} {\bibfnamefont {T.}~\bibnamefont {Ohnishi}}, \
  and\ \bibinfo {author} {\bibfnamefont {T.}~\bibnamefont {Watanabe}},\
  }\href@noop {} {\bibfield  {journal} {\bibinfo  {journal} {EPJ Data Science}\
  }\textbf {\bibinfo {volume} {5}},\ \bibinfo {pages} {1} (\bibinfo {year}
  {2016})}\BibitemShut {NoStop}%
\bibitem [{\citenamefont {Gephart}\ \emph {et~al.}(2016)\citenamefont
  {Gephart}, \citenamefont {Rovenskaya}, \citenamefont {Dieckmann},
  \citenamefont {Pace},\ and\ \citenamefont
  {Br{\"a}nnstr{\"o}m}}]{gephart2016vulnerability}%
  \BibitemOpen
  \bibfield  {author} {\bibinfo {author} {\bibfnamefont {J.~A.}\ \bibnamefont
  {Gephart}}, \bibinfo {author} {\bibfnamefont {E.}~\bibnamefont {Rovenskaya}},
  \bibinfo {author} {\bibfnamefont {U.}~\bibnamefont {Dieckmann}}, \bibinfo
  {author} {\bibfnamefont {M.~L.}\ \bibnamefont {Pace}}, \ and\ \bibinfo
  {author} {\bibfnamefont {{\AA}.}~\bibnamefont {Br{\"a}nnstr{\"o}m}},\
  }\href@noop {} {\bibfield  {journal} {\bibinfo  {journal} {Environmental
  Research Letters}\ }\textbf {\bibinfo {volume} {11}},\ \bibinfo {pages}
  {035008} (\bibinfo {year} {2016})}\BibitemShut {NoStop}%
\bibitem [{\citenamefont {Starnini}\ \emph {et~al.}(2019)\citenamefont
  {Starnini}, \citenamefont {Bogu{\~n}{\'a}},\ and\ \citenamefont
  {Serrano}}]{starnini2019interconnected}%
  \BibitemOpen
  \bibfield  {author} {\bibinfo {author} {\bibfnamefont {M.}~\bibnamefont
  {Starnini}}, \bibinfo {author} {\bibfnamefont {M.}~\bibnamefont
  {Bogu{\~n}{\'a}}}, \ and\ \bibinfo {author} {\bibfnamefont {M.~{\'A}.}\
  \bibnamefont {Serrano}},\ }\href@noop {} {\bibfield  {journal} {\bibinfo
  {journal} {Scientific Reports}\ }\textbf {\bibinfo {volume} {9}},\ \bibinfo
  {pages} {1} (\bibinfo {year} {2019})}\BibitemShut {NoStop}%
\bibitem [{\citenamefont {Klimek}\ \emph {et~al.}(2019)\citenamefont {Klimek},
  \citenamefont {Poledna},\ and\ \citenamefont
  {Thurner}}]{klimek2019quantifying}%
  \BibitemOpen
  \bibfield  {author} {\bibinfo {author} {\bibfnamefont {P.}~\bibnamefont
  {Klimek}}, \bibinfo {author} {\bibfnamefont {S.}~\bibnamefont {Poledna}}, \
  and\ \bibinfo {author} {\bibfnamefont {S.}~\bibnamefont {Thurner}},\
  }\href@noop {} {\bibfield  {journal} {\bibinfo  {journal} {Nature
  communications}\ }\textbf {\bibinfo {volume} {10}},\ \bibinfo {pages} {1}
  (\bibinfo {year} {2019})}\BibitemShut {NoStop}%
\bibitem [{\citenamefont {del Rio-Chanona}\ \emph {et~al.}(2020)\citenamefont
  {del Rio-Chanona}, \citenamefont {Korniyenko}, \citenamefont {Patnam},\ and\
  \citenamefont {Porter}}]{del2020multiplex}%
  \BibitemOpen
  \bibfield  {author} {\bibinfo {author} {\bibfnamefont {R.~M.}\ \bibnamefont
  {del Rio-Chanona}}, \bibinfo {author} {\bibfnamefont {Y.}~\bibnamefont
  {Korniyenko}}, \bibinfo {author} {\bibfnamefont {M.}~\bibnamefont {Patnam}},
  \ and\ \bibinfo {author} {\bibfnamefont {M.~A.}\ \bibnamefont {Porter}},\
  }\href@noop {} {\bibfield  {journal} {\bibinfo  {journal} {Applied Network
  Science}\ }\textbf {\bibinfo {volume} {5}},\ \bibinfo {pages} {1} (\bibinfo
  {year} {2020})}\BibitemShut {NoStop}%
\bibitem [{\citenamefont {Barrot}\ and\ \citenamefont
  {Sauvagnat}(2016)}]{barrot2016input}%
  \BibitemOpen
  \bibfield  {author} {\bibinfo {author} {\bibfnamefont {J.-N.}\ \bibnamefont
  {Barrot}}\ and\ \bibinfo {author} {\bibfnamefont {J.}~\bibnamefont
  {Sauvagnat}},\ }\href@noop {} {\bibfield  {journal} {\bibinfo  {journal} {The
  Quarterly Journal of Economics}\ }\textbf {\bibinfo {volume} {131}},\
  \bibinfo {pages} {1543} (\bibinfo {year} {2016})}\BibitemShut {NoStop}%
\bibitem [{\citenamefont {Inoue}\ and\ \citenamefont
  {Todo}(2019)}]{inoue2019firm}%
  \BibitemOpen
  \bibfield  {author} {\bibinfo {author} {\bibfnamefont {H.}~\bibnamefont
  {Inoue}}\ and\ \bibinfo {author} {\bibfnamefont {Y.}~\bibnamefont {Todo}},\
  }\href@noop {} {\bibfield  {journal} {\bibinfo  {journal} {Nature
  Sustainability}\ }\textbf {\bibinfo {volume} {2}},\ \bibinfo {pages} {841}
  (\bibinfo {year} {2019})}\BibitemShut {NoStop}%
\bibitem [{\citenamefont {Carvalho}\ \emph {et~al.}(2021)\citenamefont
  {Carvalho}, \citenamefont {Nirei}, \citenamefont {Saito},\ and\ \citenamefont
  {Tahbaz-Salehi}}]{carvalho2021supply}%
  \BibitemOpen
  \bibfield  {author} {\bibinfo {author} {\bibfnamefont {V.~M.}\ \bibnamefont
  {Carvalho}}, \bibinfo {author} {\bibfnamefont {M.}~\bibnamefont {Nirei}},
  \bibinfo {author} {\bibfnamefont {Y.~U.}\ \bibnamefont {Saito}}, \ and\
  \bibinfo {author} {\bibfnamefont {A.}~\bibnamefont {Tahbaz-Salehi}},\
  }\href@noop {} {\bibfield  {journal} {\bibinfo  {journal} {The Quarterly
  Journal of Economics}\ }\textbf {\bibinfo {volume} {136}},\ \bibinfo {pages}
  {1255} (\bibinfo {year} {2021})}\BibitemShut {NoStop}%
\bibitem [{\citenamefont {Boehm}\ \emph {et~al.}(2019)\citenamefont {Boehm},
  \citenamefont {Flaaen},\ and\ \citenamefont
  {Pandalai-Nayar}}]{boehm2019input}%
  \BibitemOpen
  \bibfield  {author} {\bibinfo {author} {\bibfnamefont {C.~E.}\ \bibnamefont
  {Boehm}}, \bibinfo {author} {\bibfnamefont {A.}~\bibnamefont {Flaaen}}, \
  and\ \bibinfo {author} {\bibfnamefont {N.}~\bibnamefont {Pandalai-Nayar}},\
  }\href@noop {} {\bibfield  {journal} {\bibinfo  {journal} {Review of
  Economics and Statistics}\ }\textbf {\bibinfo {volume} {101}},\ \bibinfo
  {pages} {60} (\bibinfo {year} {2019})}\BibitemShut {NoStop}%
\bibitem [{\citenamefont {Lee}\ \emph {et~al.}(2011)\citenamefont {Lee},
  \citenamefont {Yang}, \citenamefont {Kim}, \citenamefont {Lee}, \citenamefont
  {Goh},\ and\ \citenamefont {Kim}}]{lee2011impact}%
  \BibitemOpen
  \bibfield  {author} {\bibinfo {author} {\bibfnamefont {K.-M.}\ \bibnamefont
  {Lee}}, \bibinfo {author} {\bibfnamefont {J.-S.}\ \bibnamefont {Yang}},
  \bibinfo {author} {\bibfnamefont {G.}~\bibnamefont {Kim}}, \bibinfo {author}
  {\bibfnamefont {J.}~\bibnamefont {Lee}}, \bibinfo {author} {\bibfnamefont
  {K.-I.}\ \bibnamefont {Goh}}, \ and\ \bibinfo {author} {\bibfnamefont
  {I.-m.}\ \bibnamefont {Kim}},\ }\href@noop {} {\bibfield  {journal} {\bibinfo
   {journal} {PloS one}\ }\textbf {\bibinfo {volume} {6}},\ \bibinfo {pages}
  {e18443} (\bibinfo {year} {2011})}\BibitemShut {NoStop}%
\bibitem [{\citenamefont {Boss}\ \emph {et~al.}(2004)\citenamefont {Boss},
  \citenamefont {Elsinger}, \citenamefont {Summer},\ and\ \citenamefont
  {Thurner~4}}]{boss2004network}%
  \BibitemOpen
  \bibfield  {author} {\bibinfo {author} {\bibfnamefont {M.}~\bibnamefont
  {Boss}}, \bibinfo {author} {\bibfnamefont {H.}~\bibnamefont {Elsinger}},
  \bibinfo {author} {\bibfnamefont {M.}~\bibnamefont {Summer}}, \ and\ \bibinfo
  {author} {\bibfnamefont {S.}~\bibnamefont {Thurner~4}},\ }\href@noop {}
  {\bibfield  {journal} {\bibinfo  {journal} {Quantitative Finance}\ }\textbf
  {\bibinfo {volume} {4}},\ \bibinfo {pages} {677} (\bibinfo {year}
  {2004})}\BibitemShut {NoStop}%
\bibitem [{\citenamefont {Iori}\ \emph {et~al.}(2008)\citenamefont {Iori},
  \citenamefont {De~Masi}, \citenamefont {Precup}, \citenamefont {Gabbi},\ and\
  \citenamefont {Caldarelli}}]{iori2008network}%
  \BibitemOpen
  \bibfield  {author} {\bibinfo {author} {\bibfnamefont {G.}~\bibnamefont
  {Iori}}, \bibinfo {author} {\bibfnamefont {G.}~\bibnamefont {De~Masi}},
  \bibinfo {author} {\bibfnamefont {O.~V.}\ \bibnamefont {Precup}}, \bibinfo
  {author} {\bibfnamefont {G.}~\bibnamefont {Gabbi}}, \ and\ \bibinfo {author}
  {\bibfnamefont {G.}~\bibnamefont {Caldarelli}},\ }\href@noop {} {\bibfield
  {journal} {\bibinfo  {journal} {Journal of Economic Dynamics and Control}\
  }\textbf {\bibinfo {volume} {32}},\ \bibinfo {pages} {259} (\bibinfo {year}
  {2008})}\BibitemShut {NoStop}%
\bibitem [{\citenamefont {Battiston}\ \emph {et~al.}(2012)\citenamefont
  {Battiston}, \citenamefont {Puliga}, \citenamefont {Kaushik}, \citenamefont
  {Tasca},\ and\ \citenamefont {Caldarelli}}]{battiston2012debtrank}%
  \BibitemOpen
  \bibfield  {author} {\bibinfo {author} {\bibfnamefont {S.}~\bibnamefont
  {Battiston}}, \bibinfo {author} {\bibfnamefont {M.}~\bibnamefont {Puliga}},
  \bibinfo {author} {\bibfnamefont {R.}~\bibnamefont {Kaushik}}, \bibinfo
  {author} {\bibfnamefont {P.}~\bibnamefont {Tasca}}, \ and\ \bibinfo {author}
  {\bibfnamefont {G.}~\bibnamefont {Caldarelli}},\ }\href@noop {} {\bibfield
  {journal} {\bibinfo  {journal} {Scientific Reports}\ }\textbf {\bibinfo
  {volume} {2}},\ \bibinfo {pages} {1} (\bibinfo {year} {2012})}\BibitemShut
  {NoStop}%
\bibitem [{\citenamefont {Thurner}\ and\ \citenamefont
  {Poledna}(2013)}]{thurner2013debtrank}%
  \BibitemOpen
  \bibfield  {author} {\bibinfo {author} {\bibfnamefont {S.}~\bibnamefont
  {Thurner}}\ and\ \bibinfo {author} {\bibfnamefont {S.}~\bibnamefont
  {Poledna}},\ }\href@noop {} {\bibfield  {journal} {\bibinfo  {journal}
  {Scientific Reports}\ }\textbf {\bibinfo {volume} {3}},\ \bibinfo {pages} {1}
  (\bibinfo {year} {2013})}\BibitemShut {NoStop}%
\bibitem [{\citenamefont {Diem}\ \emph {et~al.}(2020)\citenamefont {Diem},
  \citenamefont {Pichler},\ and\ \citenamefont {Thurner}}]{diem2020minimal}%
  \BibitemOpen
  \bibfield  {author} {\bibinfo {author} {\bibfnamefont {C.}~\bibnamefont
  {Diem}}, \bibinfo {author} {\bibfnamefont {A.}~\bibnamefont {Pichler}}, \
  and\ \bibinfo {author} {\bibfnamefont {S.}~\bibnamefont {Thurner}},\ }\href
  {\doibase https://doi.org/10.1016/j.jedc.2020.103900} {\bibfield  {journal}
  {\bibinfo  {journal} {Journal of Economic Dynamics and Control}\ }\textbf
  {\bibinfo {volume} {116}},\ \bibinfo {pages} {103900} (\bibinfo {year}
  {2020})}\BibitemShut {NoStop}%
\bibitem [{\citenamefont {Pichler}\ \emph {et~al.}(2021)\citenamefont
  {Pichler}, \citenamefont {Poledna},\ and\ \citenamefont
  {Thurner}}]{pichler2021systemic}%
  \BibitemOpen
  \bibfield  {author} {\bibinfo {author} {\bibfnamefont {A.}~\bibnamefont
  {Pichler}}, \bibinfo {author} {\bibfnamefont {S.}~\bibnamefont {Poledna}}, \
  and\ \bibinfo {author} {\bibfnamefont {S.}~\bibnamefont {Thurner}},\
  }\href@noop {} {\bibfield  {journal} {\bibinfo  {journal} {Journal of
  Financial Stability}\ }\textbf {\bibinfo {volume} {52}},\ \bibinfo {pages}
  {100809} (\bibinfo {year} {2021})}\BibitemShut {NoStop}%
\bibitem [{\citenamefont {Bardoscia}\ \emph {et~al.}(2015)\citenamefont
  {Bardoscia}, \citenamefont {Battiston}, \citenamefont {Caccioli},\ and\
  \citenamefont {Caldarelli}}]{bardoscia2015debtrank}%
  \BibitemOpen
  \bibfield  {author} {\bibinfo {author} {\bibfnamefont {M.}~\bibnamefont
  {Bardoscia}}, \bibinfo {author} {\bibfnamefont {S.}~\bibnamefont
  {Battiston}}, \bibinfo {author} {\bibfnamefont {F.}~\bibnamefont {Caccioli}},
  \ and\ \bibinfo {author} {\bibfnamefont {G.}~\bibnamefont {Caldarelli}},\
  }\href@noop {} {\bibfield  {journal} {\bibinfo  {journal} {{PloS one}}\
  }\textbf {\bibinfo {volume} {10}},\ \bibinfo {pages} {e0130406} (\bibinfo
  {year} {2015})}\BibitemShut {NoStop}%
\bibitem [{\citenamefont {Poledna}\ \emph {et~al.}(2015)\citenamefont
  {Poledna}, \citenamefont {Molina-Borboa}, \citenamefont
  {Mart{\'\i}nez-Jaramillo}, \citenamefont {Van Der~Leij},\ and\ \citenamefont
  {Thurner}}]{poledna2015multi}%
  \BibitemOpen
  \bibfield  {author} {\bibinfo {author} {\bibfnamefont {S.}~\bibnamefont
  {Poledna}}, \bibinfo {author} {\bibfnamefont {J.~L.}\ \bibnamefont
  {Molina-Borboa}}, \bibinfo {author} {\bibfnamefont {S.}~\bibnamefont
  {Mart{\'\i}nez-Jaramillo}}, \bibinfo {author} {\bibfnamefont
  {M.}~\bibnamefont {Van Der~Leij}}, \ and\ \bibinfo {author} {\bibfnamefont
  {S.}~\bibnamefont {Thurner}},\ }\href@noop {} {\bibfield  {journal} {\bibinfo
   {journal} {Journal of Financial Stability}\ }\textbf {\bibinfo {volume}
  {20}},\ \bibinfo {pages} {70} (\bibinfo {year} {2015})}\BibitemShut {NoStop}%
\bibitem [{\citenamefont {Fujiwara}\ \emph {et~al.}(2016)\citenamefont
  {Fujiwara}, \citenamefont {Terai}, \citenamefont {Fujita},\ and\
  \citenamefont {Souma}}]{fujiwara2016debtrank}%
  \BibitemOpen
  \bibfield  {author} {\bibinfo {author} {\bibfnamefont {Y.}~\bibnamefont
  {Fujiwara}}, \bibinfo {author} {\bibfnamefont {M.}~\bibnamefont {Terai}},
  \bibinfo {author} {\bibfnamefont {Y.}~\bibnamefont {Fujita}}, \ and\ \bibinfo
  {author} {\bibfnamefont {W.}~\bibnamefont {Souma}},\ }\href@noop {}
  {\bibfield  {journal} {\bibinfo  {journal} {RIETI Discussion Paper Series}\
  }\textbf {\bibinfo {volume} {16-E-046}} (\bibinfo {year} {2016})}\BibitemShut
  {NoStop}%
\bibitem [{\citenamefont {Diem}\ \emph {et~al.}(2021)\citenamefont {Diem},
  \citenamefont {Borsos}, \citenamefont {Reisch}, \citenamefont {Kert{\'e}sz},\
  and\ \citenamefont {Thurner}}]{diem2021quantifying}%
  \BibitemOpen
  \bibfield  {author} {\bibinfo {author} {\bibfnamefont {C.}~\bibnamefont
  {Diem}}, \bibinfo {author} {\bibfnamefont {A.}~\bibnamefont {Borsos}},
  \bibinfo {author} {\bibfnamefont {T.}~\bibnamefont {Reisch}}, \bibinfo
  {author} {\bibfnamefont {J.}~\bibnamefont {Kert{\'e}sz}}, \ and\ \bibinfo
  {author} {\bibfnamefont {S.}~\bibnamefont {Thurner}},\ }\href@noop {}
  {\bibfield  {journal} {\bibinfo  {journal} {Available at SSRN 3826514}\ }
  (\bibinfo {year} {2021})}\BibitemShut {NoStop}%
\bibitem [{\citenamefont {Hallegatte}(2008)}]{hallegatte2008adaptive}%
  \BibitemOpen
  \bibfield  {author} {\bibinfo {author} {\bibfnamefont {S.}~\bibnamefont
  {Hallegatte}},\ }\href@noop {} {\bibfield  {journal} {\bibinfo  {journal}
  {Risk Analysis: An International Journal}\ }\textbf {\bibinfo {volume}
  {28}},\ \bibinfo {pages} {779} (\bibinfo {year} {2008})}\BibitemShut
  {NoStop}%
\bibitem [{\citenamefont {Krichene}\ \emph {et~al.}(2020)\citenamefont
  {Krichene}, \citenamefont {Inoue}, \citenamefont {Isogai},\ and\
  \citenamefont {Chakraborty}}]{krichene2020model}%
  \BibitemOpen
  \bibfield  {author} {\bibinfo {author} {\bibfnamefont {H.}~\bibnamefont
  {Krichene}}, \bibinfo {author} {\bibfnamefont {H.}~\bibnamefont {Inoue}},
  \bibinfo {author} {\bibfnamefont {T.}~\bibnamefont {Isogai}}, \ and\ \bibinfo
  {author} {\bibfnamefont {A.}~\bibnamefont {Chakraborty}},\ }\href@noop {}
  {\bibfield  {journal} {\bibinfo  {journal} {PloS one}\ }\textbf {\bibinfo
  {volume} {15}},\ \bibinfo {pages} {e0239293} (\bibinfo {year}
  {2020})}\BibitemShut {NoStop}%
\bibitem [{\citenamefont {Markhvida}\ \emph {et~al.}(2020)\citenamefont
  {Markhvida}, \citenamefont {Walsh}, \citenamefont {Hallegatte},\ and\
  \citenamefont {Baker}}]{markhvida2020quantification}%
  \BibitemOpen
  \bibfield  {author} {\bibinfo {author} {\bibfnamefont {M.}~\bibnamefont
  {Markhvida}}, \bibinfo {author} {\bibfnamefont {B.}~\bibnamefont {Walsh}},
  \bibinfo {author} {\bibfnamefont {S.}~\bibnamefont {Hallegatte}}, \ and\
  \bibinfo {author} {\bibfnamefont {J.}~\bibnamefont {Baker}},\ }\href@noop {}
  {\bibfield  {journal} {\bibinfo  {journal} {Nature Sustainability}\ }\textbf
  {\bibinfo {volume} {3}},\ \bibinfo {pages} {538} (\bibinfo {year}
  {2020})}\BibitemShut {NoStop}%
\bibitem [{\citenamefont {Haraguchi}\ and\ \citenamefont
  {Lall}(2015)}]{haraguchi2015flood}%
  \BibitemOpen
  \bibfield  {author} {\bibinfo {author} {\bibfnamefont {M.}~\bibnamefont
  {Haraguchi}}\ and\ \bibinfo {author} {\bibfnamefont {U.}~\bibnamefont
  {Lall}},\ }\href@noop {} {\bibfield  {journal} {\bibinfo  {journal}
  {International Journal of Disaster Risk Reduction}\ }\textbf {\bibinfo
  {volume} {14}},\ \bibinfo {pages} {256} (\bibinfo {year} {2015})}\BibitemShut
  {NoStop}%
\bibitem [{\citenamefont {Sweney}(2021)}]{sweney2021global}%
  \BibitemOpen
  \bibfield  {author} {\bibinfo {author} {\bibfnamefont {M.}~\bibnamefont
  {Sweney}},\ }\href
  {{https://www.theguardian.com/business/2021/mar/21/global-shortage-in-computer-chips-reaches-crisis-point}}
  {\bibfield  {journal} {\bibinfo  {journal} {{The Guardian}}\ } (\bibinfo
  {year} {2021})},\ \bibinfo {note} {{retrieved Oct. 12, 2021}}\BibitemShut
  {NoStop}%
\bibitem [{\citenamefont {Isidore}(2021)}]{isidore2021car}%
  \BibitemOpen
  \bibfield  {author} {\bibinfo {author} {\bibfnamefont {C.}~\bibnamefont
  {Isidore}},\ }\href
  {{https://edition.cnn.com/2021/10/01/business/auto-sales-third-quarter/index.html}}
  {\bibfield  {journal} {\bibinfo  {journal} {CNN Business}\ } (\bibinfo {year}
  {2021})},\ \bibinfo {note} {{retrieved Oct. 12, 2021}}\BibitemShut {NoStop}%
\bibitem [{\citenamefont {Waldersee}(2021)}]{waldersee2021chip}%
  \BibitemOpen
  \bibfield  {author} {\bibinfo {author} {\bibfnamefont {V.}~\bibnamefont
  {Waldersee}},\ }\href@noop {} {\bibfield  {journal} {\bibinfo  {journal}
  {Reuters}\ } (\bibinfo {year} {2021})},\ \bibinfo {note}
  {{https://www.reuters.com/business/autos-transportation/chip-shortage-leads-carmaker-opel-shut-german-plant-until-2022-2021-09-30/,
  retrieved Oct. 17, 2021}}\BibitemShut {NoStop}%
\bibitem [{\citenamefont {Williams}\ and\ \citenamefont
  {Bushey}(2021)}]{williams2021car}%
  \BibitemOpen
  \bibfield  {author} {\bibinfo {author} {\bibfnamefont {A.}~\bibnamefont
  {Williams}}\ and\ \bibinfo {author} {\bibfnamefont {C.}~\bibnamefont
  {Bushey}},\ }\href@noop {} {\bibfield  {journal} {\bibinfo  {journal}
  {{Financial Times}}\ } (\bibinfo {year} {2021})},\ \bibinfo {note}
  {{https://financialpost.com/financial-times/car-chip-shortage-shines-light-on-fragility-of-u-s-supply-chain,
  retrieved Oct. 17, 2021}}\BibitemShut {NoStop}%
\bibitem [{\citenamefont {Ricardo}(1891)}]{ricardo1891principles}%
  \BibitemOpen
  \bibfield  {author} {\bibinfo {author} {\bibfnamefont {D.}~\bibnamefont
  {Ricardo}},\ }\href@noop {} {\emph {\bibinfo {title} {{On the Principles of
  Political Economy and Taxation}}}}\ (\bibinfo  {publisher} {G. Bell and
  sons},\ \bibinfo {year} {1891})\BibitemShut {NoStop}%
\bibitem [{\citenamefont {Sharpe}(1964)}]{sharpe1964capital}%
  \BibitemOpen
  \bibfield  {author} {\bibinfo {author} {\bibfnamefont {W.~F.}\ \bibnamefont
  {Sharpe}},\ }\href@noop {} {\bibfield  {journal} {\bibinfo  {journal} {The
  Journal of Finance}\ }\textbf {\bibinfo {volume} {19}},\ \bibinfo {pages}
  {425} (\bibinfo {year} {1964})}\BibitemShut {NoStop}%
\bibitem [{\citenamefont {Sharpe}(1966)}]{sharpe1966mutual}%
  \BibitemOpen
  \bibfield  {author} {\bibinfo {author} {\bibfnamefont {W.~F.}\ \bibnamefont
  {Sharpe}},\ }\href@noop {} {\bibfield  {journal} {\bibinfo  {journal} {The
  Journal of Business}\ }\textbf {\bibinfo {volume} {39}},\ \bibinfo {pages}
  {119} (\bibinfo {year} {1966})}\BibitemShut {NoStop}%
\bibitem [{\citenamefont {Fama}\ and\ \citenamefont
  {French}(2015)}]{fama2015five}%
  \BibitemOpen
  \bibfield  {author} {\bibinfo {author} {\bibfnamefont {E.~F.}\ \bibnamefont
  {Fama}}\ and\ \bibinfo {author} {\bibfnamefont {K.~R.}\ \bibnamefont
  {French}},\ }\href@noop {} {\bibfield  {journal} {\bibinfo  {journal}
  {Journal of Financial Economics}\ }\textbf {\bibinfo {volume} {116}},\
  \bibinfo {pages} {1} (\bibinfo {year} {2015})}\BibitemShut {NoStop}%
\bibitem [{\citenamefont {Van~den Berg}\ and\ \citenamefont
  {Lewer}(2015)}]{van2015international}%
  \BibitemOpen
  \bibfield  {author} {\bibinfo {author} {\bibfnamefont {H.}~\bibnamefont
  {Van~den Berg}}\ and\ \bibinfo {author} {\bibfnamefont {J.~J.}\ \bibnamefont
  {Lewer}},\ }\href@noop {} {\emph {\bibinfo {title} {{International Trade and
  Economic Growth}}}}\ (\bibinfo  {publisher} {Routledge},\ \bibinfo {year}
  {2015})\BibitemShut {NoStop}%
\bibitem [{\citenamefont {Ramzan}\ \emph {et~al.}(2019)\citenamefont {Ramzan},
  \citenamefont {Sheng}, \citenamefont {Shahbaz}, \citenamefont {Song},\ and\
  \citenamefont {Jiao}}]{ramzan2019impact}%
  \BibitemOpen
  \bibfield  {author} {\bibinfo {author} {\bibfnamefont {M.}~\bibnamefont
  {Ramzan}}, \bibinfo {author} {\bibfnamefont {B.}~\bibnamefont {Sheng}},
  \bibinfo {author} {\bibfnamefont {M.}~\bibnamefont {Shahbaz}}, \bibinfo
  {author} {\bibfnamefont {J.}~\bibnamefont {Song}}, \ and\ \bibinfo {author}
  {\bibfnamefont {Z.}~\bibnamefont {Jiao}},\ }\href {\doibase
  10.1080/09638199.2019.1616805} {\bibfield  {journal} {\bibinfo  {journal}
  {The Journal of International Trade \& Economic Development}\ }\textbf
  {\bibinfo {volume} {28}},\ \bibinfo {pages} {960} (\bibinfo {year} {2019})},\
  \Eprint {http://arxiv.org/abs/https://doi.org/10.1080/09638199.2019.1616805}
  {https://doi.org/10.1080/09638199.2019.1616805} \BibitemShut {NoStop}%
\bibitem [{\citenamefont {Anderson}(2011)}]{anderson2011gravity}%
  \BibitemOpen
  \bibfield  {author} {\bibinfo {author} {\bibfnamefont {J.~E.}\ \bibnamefont
  {Anderson}},\ }\href {\doibase 10.1146/annurev-economics-111809-125114}
  {\bibfield  {journal} {\bibinfo  {journal} {Annual Review of Economics}\
  }\textbf {\bibinfo {volume} {3}},\ \bibinfo {pages} {133} (\bibinfo {year}
  {2011})},\ \Eprint
  {http://arxiv.org/abs/{https://doi.org/10.1146/annurev-economics-111809-125114}}
  {{https://doi.org/10.1146/annurev-economics-111809-125114}} \BibitemShut
  {NoStop}%
\bibitem [{\citenamefont {Goesling}\ and\ \citenamefont
  {Baker}(2008)}]{goesling2008three}%
  \BibitemOpen
  \bibfield  {author} {\bibinfo {author} {\bibfnamefont {B.}~\bibnamefont
  {Goesling}}\ and\ \bibinfo {author} {\bibfnamefont {D.~P.}\ \bibnamefont
  {Baker}},\ }\href {\doibase https://doi.org/10.1016/j.rssm.2007.11.001}
  {\bibfield  {journal} {\bibinfo  {journal} {Research in Social Stratification
  and Mobility}\ }\textbf {\bibinfo {volume} {26}},\ \bibinfo {pages} {183}
  (\bibinfo {year} {2008})}\BibitemShut {NoStop}%
\bibitem [{\citenamefont {Barro}(1999)}]{barro1999inequality}%
  \BibitemOpen
  \bibfield  {author} {\bibinfo {author} {\bibfnamefont {R.~J.}\ \bibnamefont
  {Barro}},\ }\href {\doibase 10.3386/w7038} {\emph {\bibinfo {title}
  {Inequality, Growth, and Investment}}},\ \bibinfo {type} {Working Paper}\
  \bibinfo {number} {7038}\ (\bibinfo  {institution} {National Bureau of
  Economic Research},\ \bibinfo {year} {1999})\BibitemShut {NoStop}%
\bibitem [{\citenamefont {OECD}(2015)}]{oecd2015init}%
  \BibitemOpen
  \bibfield  {author} {\bibinfo {author} {\bibnamefont {OECD}},\ }\href@noop {}
  {\enquote {\bibinfo {title} {In it together: Why less inequality benefits
  all.}}\ } (\bibinfo {year} {2015})\BibitemShut {NoStop}%
\bibitem [{\citenamefont {Turchin}(2016)}]{turchin2016ages}%
  \BibitemOpen
  \bibfield  {author} {\bibinfo {author} {\bibfnamefont {P.}~\bibnamefont
  {Turchin}},\ }\href@noop {} {\bibfield  {journal} {\bibinfo  {journal}
  {{Chaplin, CT: Beresta Books.}}\ } (\bibinfo {year} {2016})}\BibitemShut
  {NoStop}%
\bibitem [{\citenamefont {{United Nations General
  Assembly}}(2015)}]{un2015sdg}%
  \BibitemOpen
  \bibfield  {author} {\bibinfo {author} {\bibnamefont {{United Nations General
  Assembly}}},\ }\href@noop {} {\enquote {\bibinfo {title} {{Transforming Our
  World: The 2030 Agenda for Sustainable Development}},}\ } (\bibinfo {year}
  {2015}),\ \bibinfo {note} {{https://sdgs.un.org/2030agenda}}\BibitemShut
  {NoStop}%
\bibitem [{\citenamefont {DiMaggio}\ and\ \citenamefont
  {Garip}(2012)}]{dimaggio2012network}%
  \BibitemOpen
  \bibfield  {author} {\bibinfo {author} {\bibfnamefont {P.}~\bibnamefont
  {DiMaggio}}\ and\ \bibinfo {author} {\bibfnamefont {F.}~\bibnamefont
  {Garip}},\ }\href@noop {} {\bibfield  {journal} {\bibinfo  {journal} {Annual
  Review of Sociology}\ }\textbf {\bibinfo {volume} {38}},\ \bibinfo {pages}
  {93} (\bibinfo {year} {2012})}\BibitemShut {NoStop}%
\bibitem [{\citenamefont {Karimi}\ \emph {et~al.}(2018)\citenamefont {Karimi},
  \citenamefont {G{\'e}nois}, \citenamefont {Wagner}, \citenamefont {Singer},\
  and\ \citenamefont {Strohmaier}}]{karimi2018homophily}%
  \BibitemOpen
  \bibfield  {author} {\bibinfo {author} {\bibfnamefont {F.}~\bibnamefont
  {Karimi}}, \bibinfo {author} {\bibfnamefont {M.}~\bibnamefont {G{\'e}nois}},
  \bibinfo {author} {\bibfnamefont {C.}~\bibnamefont {Wagner}}, \bibinfo
  {author} {\bibfnamefont {P.}~\bibnamefont {Singer}}, \ and\ \bibinfo {author}
  {\bibfnamefont {M.}~\bibnamefont {Strohmaier}},\ }\href@noop {} {\bibfield
  {journal} {\bibinfo  {journal} {Scientific Reports}\ }\textbf {\bibinfo
  {volume} {8}},\ \bibinfo {pages} {1} (\bibinfo {year} {2018})}\BibitemShut
  {NoStop}%
\bibitem [{\citenamefont {Leduc}\ and\ \citenamefont
  {Thurner}(2017)}]{leduc2017incentivizing}%
  \BibitemOpen
  \bibfield  {author} {\bibinfo {author} {\bibfnamefont {M.~V.}\ \bibnamefont
  {Leduc}}\ and\ \bibinfo {author} {\bibfnamefont {S.}~\bibnamefont
  {Thurner}},\ }\href@noop {} {\bibfield  {journal} {\bibinfo  {journal}
  {Journal of Economic Dynamics and Control}\ }\textbf {\bibinfo {volume}
  {82}},\ \bibinfo {pages} {44} (\bibinfo {year} {2017})}\BibitemShut {NoStop}%
\bibitem [{\citenamefont {Poledna}\ and\ \citenamefont
  {Thurner}(2016)}]{poledna2016elimination}%
  \BibitemOpen
  \bibfield  {author} {\bibinfo {author} {\bibfnamefont {S.}~\bibnamefont
  {Poledna}}\ and\ \bibinfo {author} {\bibfnamefont {S.}~\bibnamefont
  {Thurner}},\ }\href@noop {} {\bibfield  {journal} {\bibinfo  {journal}
  {Quantitative Finance}\ }\textbf {\bibinfo {volume} {16}},\ \bibinfo {pages}
  {1599} (\bibinfo {year} {2016})}\BibitemShut {NoStop}%
\bibitem [{\citenamefont {Chakraborty}\ and\ \citenamefont
  {Ikeda}(2020)}]{chakraborty2020testing}%
  \BibitemOpen
  \bibfield  {author} {\bibinfo {author} {\bibfnamefont {A.}~\bibnamefont
  {Chakraborty}}\ and\ \bibinfo {author} {\bibfnamefont {Y.}~\bibnamefont
  {Ikeda}},\ }\href@noop {} {\bibfield  {journal} {\bibinfo  {journal} {PloS
  one}\ }\textbf {\bibinfo {volume} {15}},\ \bibinfo {pages} {e0239669}
  (\bibinfo {year} {2020})}\BibitemShut {NoStop}%
\end{thebibliography}%
	
	\section*{Acknowledgements}
	We thank Y. Ikeda for providing the S\&P Capital IQ data. 
	The project was supported by Austrian Science Fund FWF under I 3073-N32, Austrian Science Promotion Agency FFG under 857136, and 
	Hochschuljubil\"aumsstiftung of the Austrian National Bank OeNB under P17795 2018-2021.

	\section*{Author contributions statement}
	All authors conceived and designed the study.
	A.C. analyzed the data.
	All authors discussed the results and contributed to the manuscript.
	A.C., T.R. and S.T. wrote the paper.
	
	\section*{Additional information}
	\textbf{Competing interests} The authors declare no competing interests.

	\newpage
	\onecolumngrid  
	
	\FloatBarrier
	\section*{Supplementary Information}
	
	\renewcommand{\figurename}{SI Fig.}
	 \setcounter{table}{0}
        \renewcommand{\thetable}{S\arabic{table}}%
        \setcounter{figure}{0}
        \renewcommand{\thefigure}{S\arabic{figure}}%
	
	\subsection*{SI Text 1: DebtRank}
	The DebtRank $R_i$ of a firm in a supply network \cite{fujiwara2016debtrank} describes the overall reduction in economic activity subsequent to the default of an initial firm $i$ and the cascade of defaults caused by it. Here, we provide a brief review of the underlying cascading mechanism and show how to calculate the indicator $R_i$ as used in \cite{fujiwara2016debtrank}.
	
	Let us consider a directed and weighted network with nodes $i = 1, 2, ... ,N$. A link from node $j$ to $i$ has a weight $w_{ji} \in [0,1]$ that represents a relative dependency of $i$ to $j$. At any time-step t, the nodes are characterised by their state $S_i(t)$ and the amount of financial distress $h_i(t) \in [0,1]$. The state of a node $S_i(t) \in \{A,D,I\}$ can be  ``Active'', ``Distressed'' or ``Inactive'' at time $t$.
	We start with the following initial configurations: 
	\begin{equation*}
		h_i(0)=
		\begin{cases}
			1, & \text{if}\ i \in \cal{M} \\
			0, & \text{otherwise}
		\end{cases}
	\end{equation*}
	and 
	\begin{equation*}
		S_i(0)=
		\begin{cases}
			D, & \text{if}\ i \in \cal{M} \\
			A, & \text{otherwise}
		\end{cases}
	\end{equation*}
	where $\cal{M}$ is an initial set of ``Distressed'' nodes, which can be a single node as well. 
	We update the amount of distress as 
	$$h_i(t) = min\Bigg[1, ~ ~ h_i(t-1) + \sum_{j:S_j(t-1)=D} w_{ji} h_j(t-1)\Bigg], $$
	where the summation is taken over all the neighbours of $i$ having state $S_j(t-1) = D$. 
	We also update the state of each node simultaneously as follows 
	\begin{equation*}
		S_i(t)=
		\begin{cases}
			D, & \text{if}\ h_i > 0 \ \text{and} \ S_i(t-1)=A \\
			I, & \text{if}\ S_i(t-1)=D \\
			S_i(t-1), & \text{otherwise}
		\end{cases}
	\end{equation*}
	Note that at the next time step, a node in state $D$ becomes $I$, which does not propagate any distress to others afterwards. This helps to exclude, in the case of cycles, an infinite number of repercussions in shock propagation. However, an $I$ node continues to receive distress from its distressed neighbours without affecting others. 
	The propagation terminates after a finite number of time steps $T$. In the following we set $\mathcal{M} = \{i\} $ to just consider the default of one node, the generalization to a set of nodes is straightforward. 
	We use the matrix $D$ with element $D_{ij} = h_{j}(T)$ denoting the distress firm $j$ receives if firm $i$ defaults (the loss in economic production firm $j$ experiences if firm $i$ defaults). The column $j$ contains the shock firm $j$ is exposed to. Row $i$ contains the shock firm $i$ causes to all other firms $j$ when it defaults. 
	The total amount of distress, i.e. the total loss of production, in the system due to the initially distressed node $i$ is measured 
	as its DebtRank $$R_i = \frac{\sum_j D_{ij} q_j}{\sum_j q_j} ,$$ where $q_i$ is the size of the nodes $i$. We include the effect of the initial set of distressed nodes. The quantities discussed in the main text are based on different aggregations of $D_{ij}$.
	
	Since our global supply chain network does not have link-weights and node size information, we have assigned each edge equal weight $w_{ji}=1/k_i^{in}$ and associate the node-size with it's degree $q_i=k_i$, where  $k_i$ and $ k_i^{in}$ represent degree and in-degree of the $i$-th node, respectively. We choose degree $k_i$ as size proxy, to be consistent and self-contained in the supply network dataset. However, one could also use other size proxies such as value added, turnover or employees.

	\subsection*{SI Text 2: Toy example of a cascade of production interruptions}
\begin{figure}[!h]
		\centering
		\includegraphics[width=0.33\linewidth]{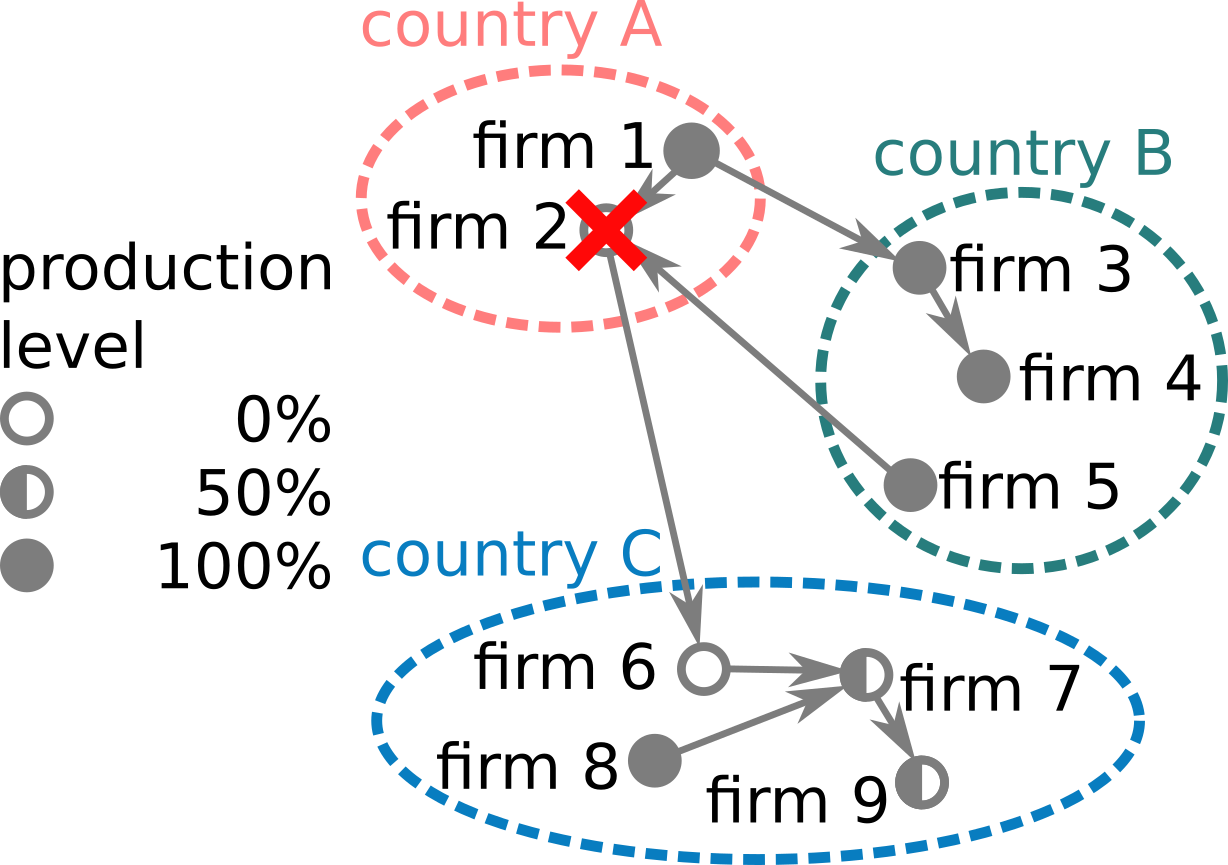}
		\caption{Visualization of a cascade of production reductions, in a toy economy of three nations and nine firms, subsequent to the default of firm 2, marked by a red cross. The filling of the nodes (pie chart) indicates their remaining production level after the shock propagation. The production of a firm is reduced proportional to the reduction in trade at the supplier divided by the number of suppliers the firm has.}
		\label{fig:SI_toyexample_firm2}
	\end{figure}

    In SI Fig.~\ref{fig:SI_toyexample_firm2} we show the cascade subsequent to the default of firm 2. Compared to the default of firm 1, country C is hit harder and country B is not affected at all.
    
	\subsection*{SI Text 3: Detailed exposure matrix}

\begin{figure}[htbp]
		\includegraphics[width=\linewidth]{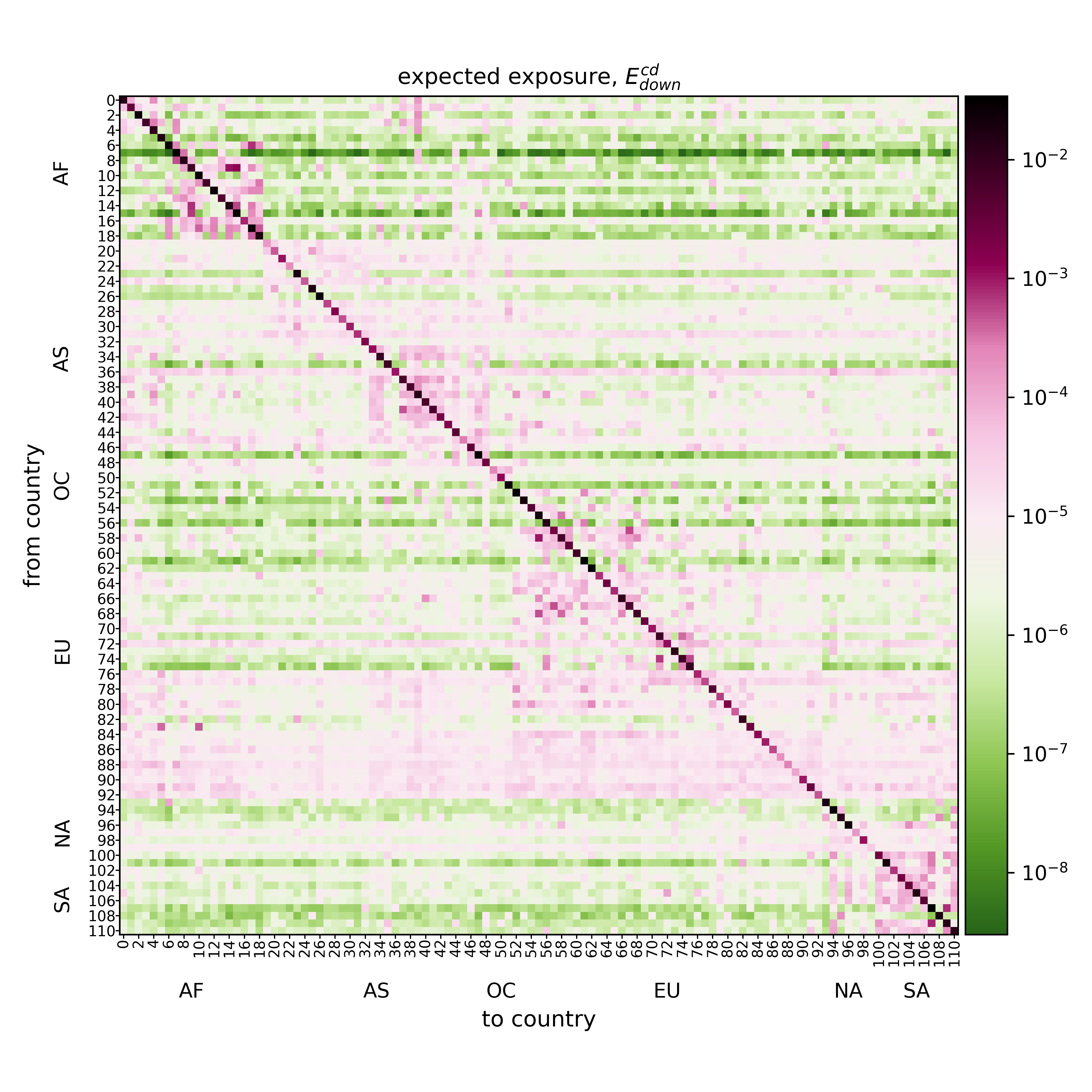}
		\vspace{-8mm}
		\caption{The network structure of international downstream exposure. Expected exposure $E^{cd}_{down}$ sorted by continent and region on a logarithmic scale. The continents are sorted as follows: Africa (AF), Asia (AS), Oceania and Australia (OC), European (EU), North America (NA), and South America (SA). The numbering corresponds to the index in SI Tab. \ref{tab:SI:country_names}, where the countries' name and region are listed.}
		\label{fig:heatmap_large}
	\end{figure}

	In the main text we show that the expected downstream exposure between countries, $E^{cd}_{down}$, is clearly structured by geographic regions. Exposures are highest within countries, then within regions and continents. In SI Fig.~\ref{fig:heatmap_large} we show a large version of the same plot, such that we can label the individual countries. The numbers correspond to the index column in SI Tab.~\ref{tab:SI:country_names}.
	
	In Africa (AF) we find two blocks corresponding to north African and sub-Sahara countries. Most of their high exposures are within their region, but it is also instructive to consider the columns of the matrix, showing that, e.g. some north African countries are strongly exposured to Asia and Europe, but not to the Americas.
	
	In Asia (AS) we have three regions, East Asia and Pacific, Middle East and South Asia, but observe only two prominent blocks, suggesting that the regional classification does not fully represent economic regions. Several Asian countries expose countries around the world to medium exposure, highlighted by horizontal white lines.
	
	Oceania and Australia (OC) does not show up as a prominent block, but is rather connected to South Asia. Further, as an industrialized region it creates exposure to most other countries in the world.
	
	The European (EU) countries ares split into four groups, Central and Eastern Europe (CEE), Northern Europe (NE), Southern Europe (SE), and Western Europe (WE). CEE and NE countries form relatively well separated blocks, while SE and WE highly expose all countries. This strongly suggests that SE and WE countries are integrated into the world market more strongly than NE and CEE countries.
	
	In North America (NA) there is no clear block structure visible. However, for Canada (index 97), Mexico (98) and the USA (99) the horizontal white lines show that they highly expose most countries around the globe. The magenta vertical line for the USA between row 100 and 110 suggests that the USA is also highly exposed to disruptions in Southern America.
	
	South America (SA) is visible as a very prominent, dark magenta block. This implies that the economies in SA are tightly interwoven and expose each other strongly.
	
	\begin{longtable}{lllll}
		\caption{
			Geographical grouping of countries into continents, regions and income groups.
			\label{tab:SI:country_names}
		}
		\\
		index &                       country name &      continent &                        region &   income group\\
		
		\hline
		\endfirsthead 
		\caption{Continuation:}\\
		index &                       country name &      continent &                        region &   income group\\
		
		\hline
		\endhead 
		\endfoot
		\hline
		
		\endlastfoot
		\toprule
		
		0   &                            Algeria &                 Africa &  Middle East and North Africa &          low \\
		1   &                              Egypt &                 Africa &  Middle East and North Africa &          low \\
		2   &                              Libya &                 Africa &  Middle East and North Africa &       middle \\
		3   &                            Morocco &                 Africa &  Middle East and North Africa &          low \\
		4   &                            Tunisia &                 Africa &  Middle East and North Africa &          low \\
		5   &                             Angola &                 Africa &            Sub-Saharan Africa &          low \\
		6   &                           Botswana &                 Africa &            Sub-Saharan Africa &       middle \\
		7   &                      Côte d'Ivoire &                 Africa &            Sub-Saharan Africa &          low \\
		8   &                              Ghana &                 Africa &            Sub-Saharan Africa &          low \\
		9   &                              Kenya &                 Africa &            Sub-Saharan Africa &          low \\
		10  &                         Mozambique &                 Africa &            Sub-Saharan Africa &          low \\
		11  &                          Mauritius &                 Africa &            Sub-Saharan Africa &       middle \\
		12  &                            Namibia &                 Africa &            Sub-Saharan Africa &          low \\
		13  &                            Nigeria &                 Africa &            Sub-Saharan Africa &          low \\
		14  &       Tanzania, United Republic of &                 Africa &            Sub-Saharan Africa &          low \\
		15  &                             Uganda &                 Africa &            Sub-Saharan Africa &          low \\
		16  &                       South Africa &                 Africa &            Sub-Saharan Africa &       middle \\
		17  &                             Zambia &                 Africa &            Sub-Saharan Africa &          low \\
		18  &                           Zimbabwe &                 Africa &            Sub-Saharan Africa &          low \\
		19  &                              China &                   Asia &         East Asia and Pacific &       middle \\
		20  &                          Hong Kong &                   Asia &         East Asia and Pacific &         high \\
		21  &                          Indonesia &                   Asia &         East Asia and Pacific &          low \\
		22  &                              Japan &                   Asia &         East Asia and Pacific &         high \\
		23  &                           Cambodia &                   Asia &         East Asia and Pacific &          low \\
		24  &                 Korea, Republic of &                   Asia &         East Asia and Pacific &         high \\
		25  &                              Macao &                   Asia &         East Asia and Pacific &         high \\
		26  &                           Mongolia &                   Asia &         East Asia and Pacific &          low \\
		27  &                           Malaysia &                   Asia &         East Asia and Pacific &       middle \\
		28  &                        Philippines &                   Asia &         East Asia and Pacific &          low \\
		29  &                          Singapore &                   Asia &         East Asia and Pacific &         high \\
		30  &                           Thailand &                   Asia &         East Asia and Pacific &       middle \\
		31  &                             Taiwan &                   Asia &         East Asia and Pacific &         high \\
		32  &                           Viet Nam &                   Asia &         East Asia and Pacific &          low \\
		33  &               United Arab Emirates &                   Asia &  Middle East and North Africa &         high \\
		34  &                            Bahrain &                   Asia &  Middle East and North Africa &         high \\
		35  &          Iran, Islamic Republic of &                   Asia &  Middle East and North Africa &          low \\
		36  &                             Israel &                   Asia &  Middle East and North Africa &         high \\
		37  &                             Jordan &                   Asia &  Middle East and North Africa &          low \\
		38  &                             Kuwait &                   Asia &  Middle East and North Africa &         high \\
		39  &                            Lebanon &                   Asia &  Middle East and North Africa &       middle \\
		40  &                               Oman &                   Asia &  Middle East and North Africa &       middle \\
		41  &                              Qatar &                   Asia &  Middle East and North Africa &         high \\
		42  &                       Saudi Arabia &                   Asia &  Middle East and North Africa &         high \\
		43  &                             Turkey &                   Asia &  Middle East and North Africa &       middle \\
		44  &                         Bangladesh &                   Asia &                    South Asia &          low \\
		45  &                              India &                   Asia &                    South Asia &          low \\
		46  &                          Sri Lanka &                   Asia &                    South Asia &          low \\
		47  &                              Nepal &                   Asia &                    South Asia &          low \\
		48  &                           Pakistan &                   Asia &                    South Asia &          low \\
		49  &                          Australia &  Australia and Oceania &         East Asia and Pacific &         high \\
		50  &                        New Zealand &  Australia and Oceania &         East Asia and Pacific &         high \\
		51  &                   Papua New Guinea &  Australia and Oceania &         East Asia and Pacific &          low \\
		52  &                            Armenia &                 Europe &    Central and Eastern Europe &          low \\
		53  &                         Azerbaijan &                 Europe &    Central and Eastern Europe &          low \\
		54  &                           Bulgaria &                 Europe &    Central and Eastern Europe &       middle \\
		55  &             Bosnia and Herzegovina &                 Europe &    Central and Eastern Europe &          low \\
		56  &                            Belarus &                 Europe &    Central and Eastern Europe &       middle \\
		57  &                            Czechia &                 Europe &    Central and Eastern Europe &         high \\
		58  &                            Croatia &                 Europe &    Central and Eastern Europe &       middle \\
		59  &                            Hungary &                 Europe &    Central and Eastern Europe &       middle \\
		60  &                         Kazakhstan &                 Europe &    Central and Eastern Europe &       middle \\
		61  &               Moldova, Republic of &                 Europe &    Central and Eastern Europe &          low \\
		62  &                    North Macedonia &                 Europe &    Central and Eastern Europe &       middle \\
		63  &                             Poland &                 Europe &    Central and Eastern Europe &       middle \\
		64  &                            Romania &                 Europe &    Central and Eastern Europe &       middle \\
		65  &                 Russian Federation &                 Europe &    Central and Eastern Europe &       middle \\
		66  &                             Serbia &                 Europe &    Central and Eastern Europe &       middle \\
		67  &                           Slovakia &                 Europe &    Central and Eastern Europe &       middle \\
		68  &                           Slovenia &                 Europe &    Central and Eastern Europe &         high \\
		69  &                            Ukraine &                 Europe &    Central and Eastern Europe &          low \\
		70  &                            Denmark &                 Europe &               Northern Europe &         high \\
		71  &                            Estonia &                 Europe &               Northern Europe &       middle \\
		72  &                            Finland &                 Europe &               Northern Europe &         high \\
		73  &                            Iceland &                 Europe &               Northern Europe &         high \\
		74  &                          Lithuania &                 Europe &               Northern Europe &       middle \\
		75  &                             Latvia &                 Europe &               Northern Europe &       middle \\
		76  &                             Norway &                 Europe &               Northern Europe &         high \\
		77  &                             Sweden &                 Europe &               Northern Europe &         high \\
		78  &                             Cyprus &                 Europe &               Southern Europe &         high \\
		79  &                              Spain &                 Europe &               Southern Europe &         high \\
		80  &                             Greece &                 Europe &               Southern Europe &       middle \\
		81  &                              Italy &                 Europe &               Southern Europe &         high \\
		82  &                              Malta &                 Europe &               Southern Europe &         high \\
		83  &                           Portugal &                 Europe &               Southern Europe &         high \\
		84  &                            Austria &                 Europe &                Western Europe &         high \\
		85  &                            Belgium &                 Europe &                Western Europe &         high \\
		86  &                        Switzerland &                 Europe &                Western Europe &         high \\
		87  &                            Germany &                 Europe &                Western Europe &         high \\
		88  &                             France &                 Europe &                Western Europe &         high \\
		89  &                     United Kingdom &                 Europe &                Western Europe &         high \\
		90  &                            Ireland &                 Europe &                Western Europe &         high \\
		91  &                         Luxembourg &                 Europe &                Western Europe &         high \\
		92  &                        Netherlands &                 Europe &                Western Europe &         high \\
		93  &                            Bahamas &          North America &   Latin America and Caribbean &         high \\
		94  &                 Dominican Republic &          North America &   Latin America and Caribbean &       middle \\
		95  &                            Jamaica &          North America &   Latin America and Caribbean &          low \\
		96  &                             Panama &          North America &   Latin America and Caribbean &       middle \\
		97  &                             Canada &          North America &                 North America &         high \\
		98  &                             Mexico &          North America &                 North America &       middle \\
		99  &                      United States &          North America &                 North America &         high \\
		100 &                          Argentina &          South America &   Latin America and Caribbean &       middle \\
		101 &    Bolivia, Plurinational State of &          South America &   Latin America and Caribbean &          low \\
		102 &                             Brazil &          South America &   Latin America and Caribbean &       middle \\
		103 &                              Chile &          South America &   Latin America and Caribbean &       middle \\
		104 &                           Colombia &          South America &   Latin America and Caribbean &       middle \\
		105 &                            Ecuador &          South America &   Latin America and Caribbean &       middle \\
		106 &                               Peru &          South America &   Latin America and Caribbean &       middle \\
		107 &                           Paraguay &          South America &   Latin America and Caribbean &          low \\
		108 &                Trinidad and Tobago &          South America &   Latin America and Caribbean &       middle \\
		109 &                            Uruguay &          South America &   Latin America and Caribbean &       middle \\
		110 &  Venezuela, Bolivarian Republic of &          South America &   Latin America and Caribbean &          low \\
	\end{longtable}
	
	\subsection*{SI Text 4: Correlation of direct links and country country exposure}

\begin{figure}[ht]
		\centering
		\includegraphics[width=0.5\linewidth]{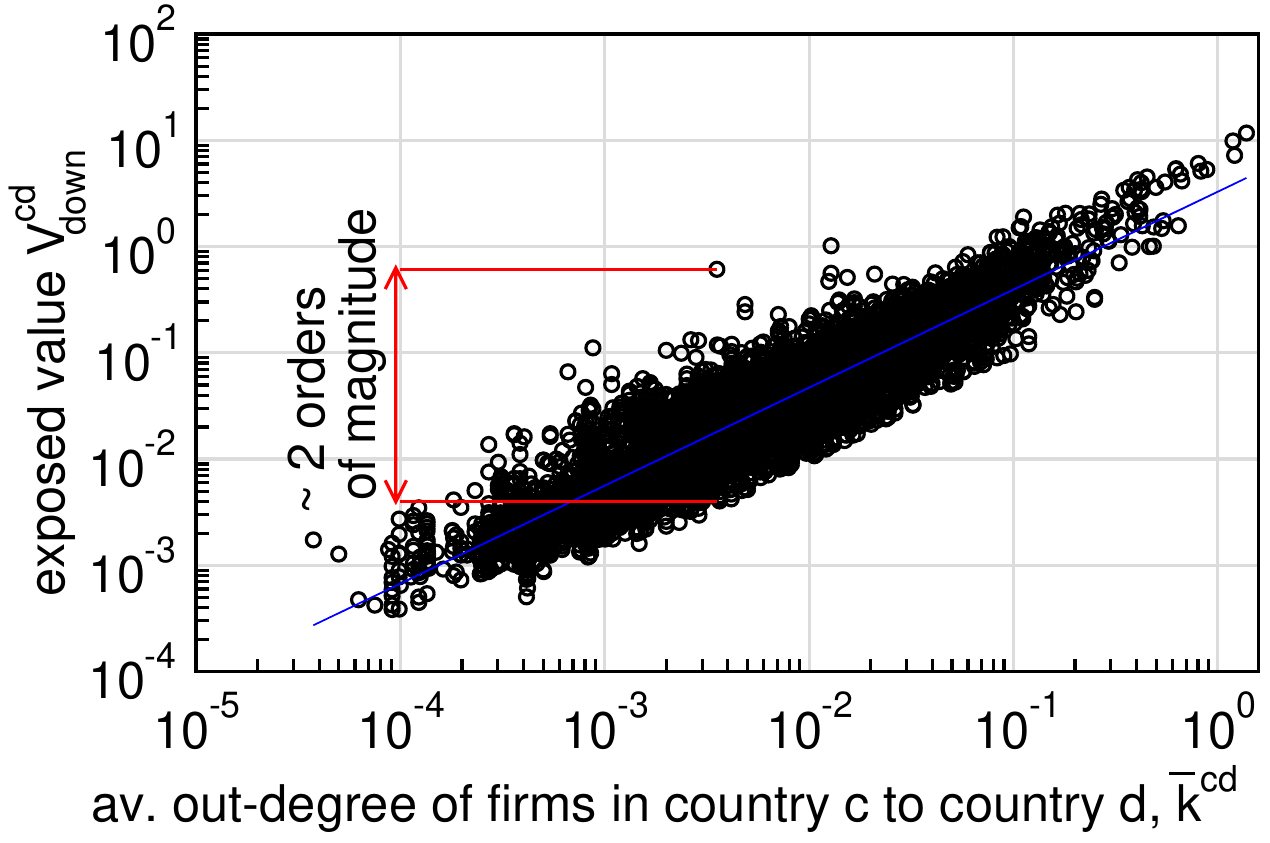}
		\caption{Exposed economic value $V^{cd}_{down}$ is plotted against the average number of out-links of firms in country $c$ to country $d$, $\bar{k}^{cd}$. The blue line represents the regression line: $V^{cd}_{down} \sim \bar{k}^{cd~0.92}$ estimated by a log-log least square fit. Although $V^{cd}_{down}$ correlates strongly with $\bar{k}^{cd}$, for a given level of connectedness, $\bar{k}^{cd}$, the exposed value, $V^{cd}_{down}$, ranges over two orders of magnitude.}
		\label{fig:d_vs_DRcd}
	\end{figure}

	Direct trade links represent first order (i.e. direct) exposures. Consequently, the \emph{exposed value} in country $c$ subsequent to the default of a firm in country $d$, $V^{cd}$, increases as function of the average number of links firms in country $c$ have to country $d$. 
	We quantify the direct influence of one country on another by the average number of out-links of a firm in country $c$ to the firms in country $d$, $\bar{k}^{cd} = A^{cd} / N^c$. 
	In Fig.~\ref{fig:d_vs_DRcd} we plot $V^{cd}$ on the y-axis and $\bar{k}^{cd}$  on the x-axis. We find that $V^{cd}$ spans five orders of magnitude. Although we observe a high correlation between $\bar{k}^{cd}$ and $V^{cd}$ (Pearson's $r = 0.93$ $p <  10^{-15}$),  we find large variations in $V^{cd}$, for a given level of average outlinks. The red arrow in SI Fig.~\ref{fig:d_vs_DRcd} highlights that this variation can be more than two orders of magnitude for a given value of average out-degree. This variation can be attributed to higher order exposures that are heavily influenced by the network topology. For a fully connected network --equivalently to an aggregated version of the network-- $V^{cd}$ and $\bar{k}^{cd}$ are perfectly correlated.\\
	
	\subsection*{SI Text 5: Multilinear regression model for GDP per capita}
	We perform the following multi linear regression models for GDP per capita and report the results in SI Tab.~\ref{table:1}. \\
	
	Model 1: GDP/capita $\sim $ Intercept + Total downstream imported distress + GDP + Exports + Imports + Exports/capita +Import/capita\\
	
	The model explains 67\% of variance (adjusted $R^2 = 0.67$) and the significant covariates are $E^d_{down}$ and exports per capita.
	We perform a second regression with only $E^d_{down}$ and exports per capita as covariates which explains a comparable amount of variance (adjusted $R^2=0.65$, $p< 10^{-15}$), see SI Tab.~\ref{table:1}. \\
	
	Model 2: GDP/capita $\sim $ Intercept + Total downstream imported distress + Exports/capita 
	
	\begin{table}[h!]
		\caption{ Multi linear regression table for GDP/capita.}
		\centering
		\begin{tabular}{|l c | c |} 
			\hline
			Variables & Model 1 & Model 2 \\ 
			\hline
			Intercept & $2.067 \times 10^{4}$  *** & $2.503 \times 10^{4}$  ***\\ 
			Total downstream imported distress, $E^d_{down}$ & $-8.149\times 10^{6}$ ** & $-1.081\times 10^{7}$ *** \\
			GDP & $-8.371\times 10^{-10}$ & \\
			Exports  & $-1.707\times10^{-8}$ & \\
			Imports &  $2.974 \times 10^{-8}$ &  \\  
			Exports/capita & $1.045$   ** & $0.596$ *** \\
			Imports/capita & $-0.556$ & \\
			\hline
			Observations & $105$  & $105$\\
			Adjusted R2 & $0.67$ & $0.65$\\
			p value &  $< 2.2\times 10^{-16}$ & $< 2.2\times 10^{-16}$\\
			\hline
		\end{tabular}
		\begin{flushleft}
Significance codes for p value: $0 \leq *** \leq 0.001 \leq ** \leq 0.01 \leq * \leq 0.05$
  \end{flushleft}
		\label{table:1}
	\end{table}
	
	\subsection*{SI Text 6: GDP growth vs. total exposure $E^d$}
\begin{figure}[htbp]
		\centering
		\includegraphics[width=0.5\linewidth]{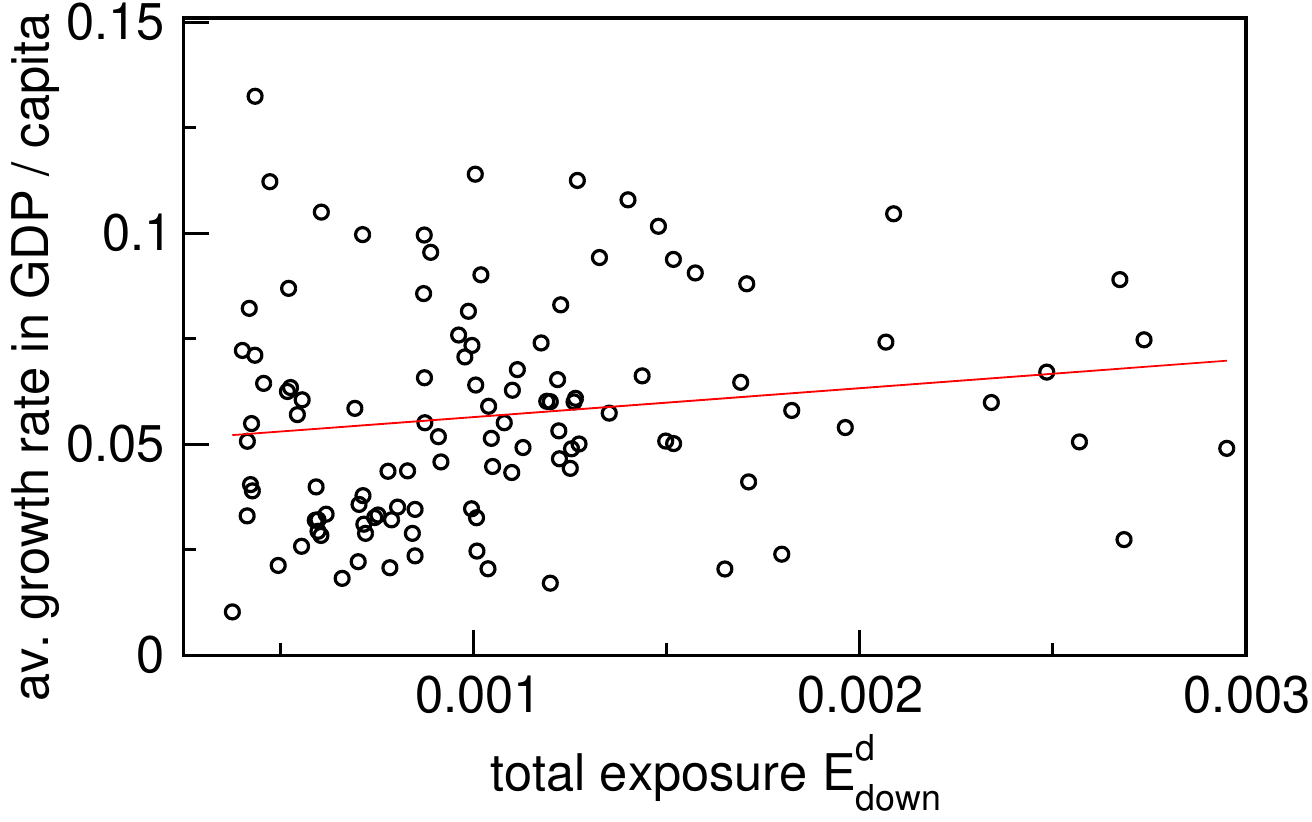}
		\caption{Average growth in GDP per capita over 20 years is plotted against total exposure $E^d_{down}$. The red line represents the  best  linear fit to  the  data  of  form $y = 0.05 + 6.85 x$. There is no significant linear relationship between the two variables.}
		\label{fig:SI:growthvsexpsure}
	\end{figure}

	Intuitively we expect higher gains from higher risks. For lack of a better indicator, in SI Fig.~\ref{fig:SI:growthvsexpsure} we compare total exposure $E^d_{down}$ with average annual growth of GDP per capita. International trade is generally associated with higher GDP growth~\cite{van2015international,ramzan2019impact}. We find no significant correlation between $E^d_{down}$ and the average GDP per capita growth rate over 20 years, between 1998 and 2018 (Pearson $r=0.15, p = 0.12 $), see SI Fig. \ref{fig:SI:growthvsexpsure}. We test our results for robustness by calculating the correlation with 5- and 10-year growth, but find no correlations (10-year growth: Pearson $r=0.11, p = 0.26 $; 5-year growth: Pearson $r=-0.03, p = 0.73 $). Note that we are limited by data availability and can only compare exposures on the 2017 network with historical growth rates.
	
	\subsection*{SI Text 7: Upstream exposures}
\begin{figure}[ht]
		\centering
		\includegraphics[width=0.5\linewidth]{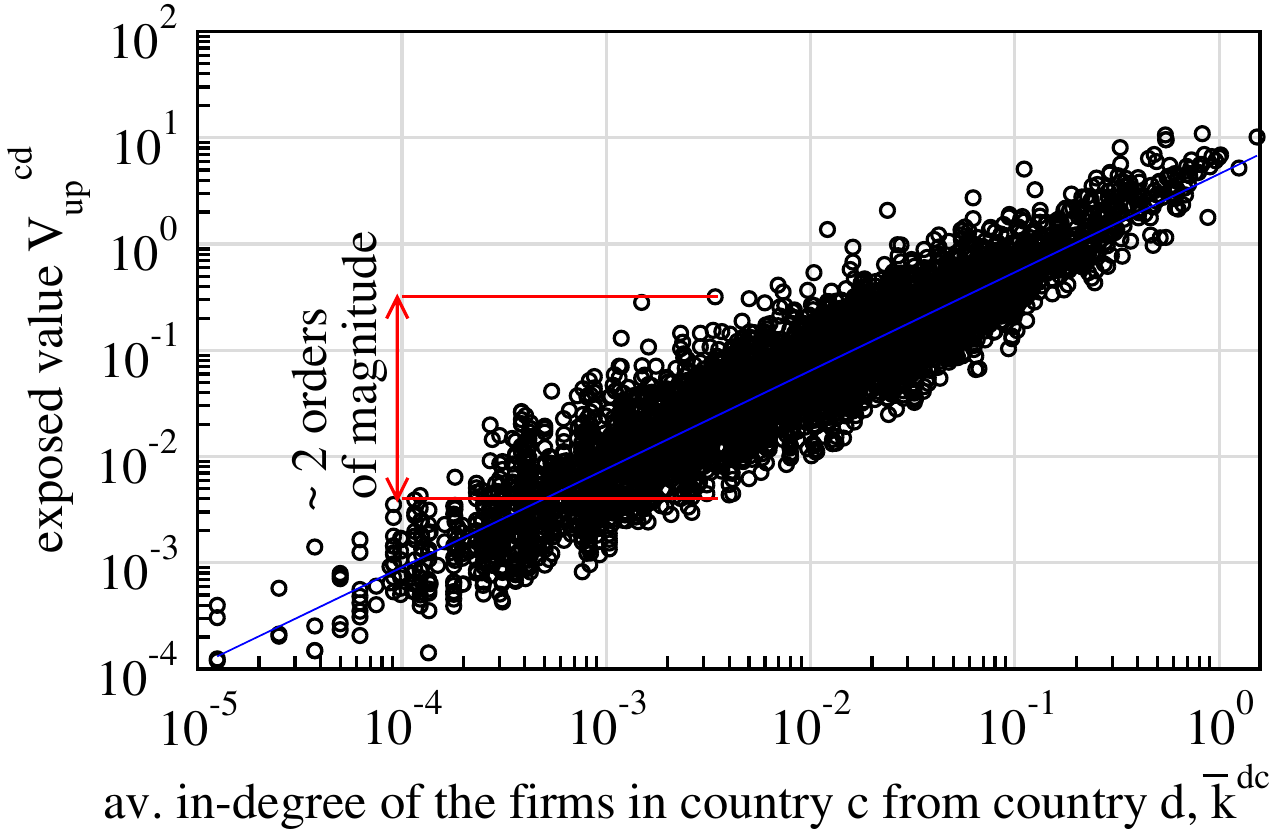}
		\caption{The Exposed economic value for upstream  $V^{cd}_{up}$ is plotted with average in-degree of firms in country $c$ from firms in country $d$, $\overline{k}^{dc}$. The blue line represents the best power-law fit to the data of form: $V^{cd}_{up} \sim (\bar{k}^{dc})^{0.93}$.}
		\label{fig:d_vs_DRcd_up}
	\end{figure}

	In the main text, we have studied the downstream propagation of shocks, reflecting the impact of a supplier default. We can also study upstream cascades to show the impact of a customer's default. 
	SI Figure~\ref{fig:d_vs_DRcd_up} shows the variation of upstream exposed economic value $V^{cd}_{up}$ with average in-degree of firms in country $c$ from the firms in country $d$, $\bar{k}^{dc}$. Here also, we observe that these two quantities are strongly correlated ($r=0.88, p<10^{-15}$), but retain a lot of variance, indicated by a red arrow which shows a variation of two orders of magnitude for a given value of $\bar{k}^{dc}$. 
	
\begin{figure}[ht]
	    \centering
		\includegraphics[width=0.75\linewidth]{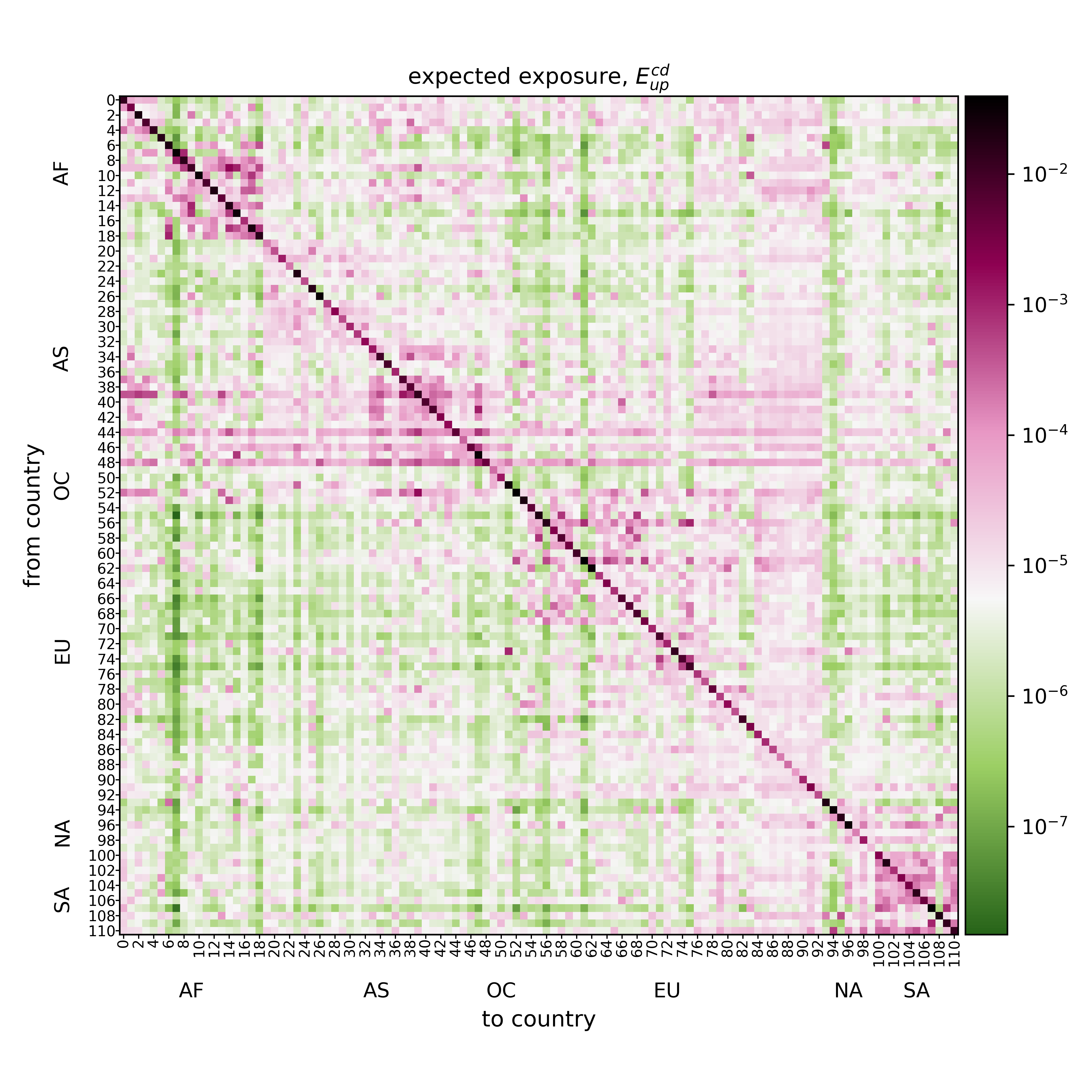}
		\caption{The bilateral upstream distress ($\log$-scale) for all pairs of countries is shown as a heat map. The countries are grouped according to their regional classification. The country names and their regional classification can be found in SI Tab.~\ref{tab:SI:country_names}. The matrix resembles a transposed $E^{cd}_{down}$ in terms of the block structure and the exposures of western Europe, but also highlights some differences. For instance, several Asian countries create high upstream exposures, such as Lebanon (index 19), Bangladesh (44), Sri Lanka (46) and Pakistan (48).} 
		\label{fig:heatmap_up}
	\end{figure}

	We explore the structure of the upstream country-country exposure matrix $E^{cd}_{up}$ in SI Fig.~\ref{fig:heatmap_up}.
	Contrary to intuition, it is not simply the transpose of its downstream counterpart. It shows similar structures, such as high values within countries and similar clusters of geographic regions. Confirming expectation, some countries that created a lot of downstream exposure, such the block of northern, southern and western European countries, now receive downstream distress when passing buyer-supplier relations the opposite way. However, some countries, such as several Middle Eastern countries, very prominently create a lot of upstream exposure that was not visible in the downstream cascade, see SI Fig.~\ref{fig:heatmap_up}.
	
\begin{figure}[ht]
		\centering
		\includegraphics[width=0.66\linewidth]{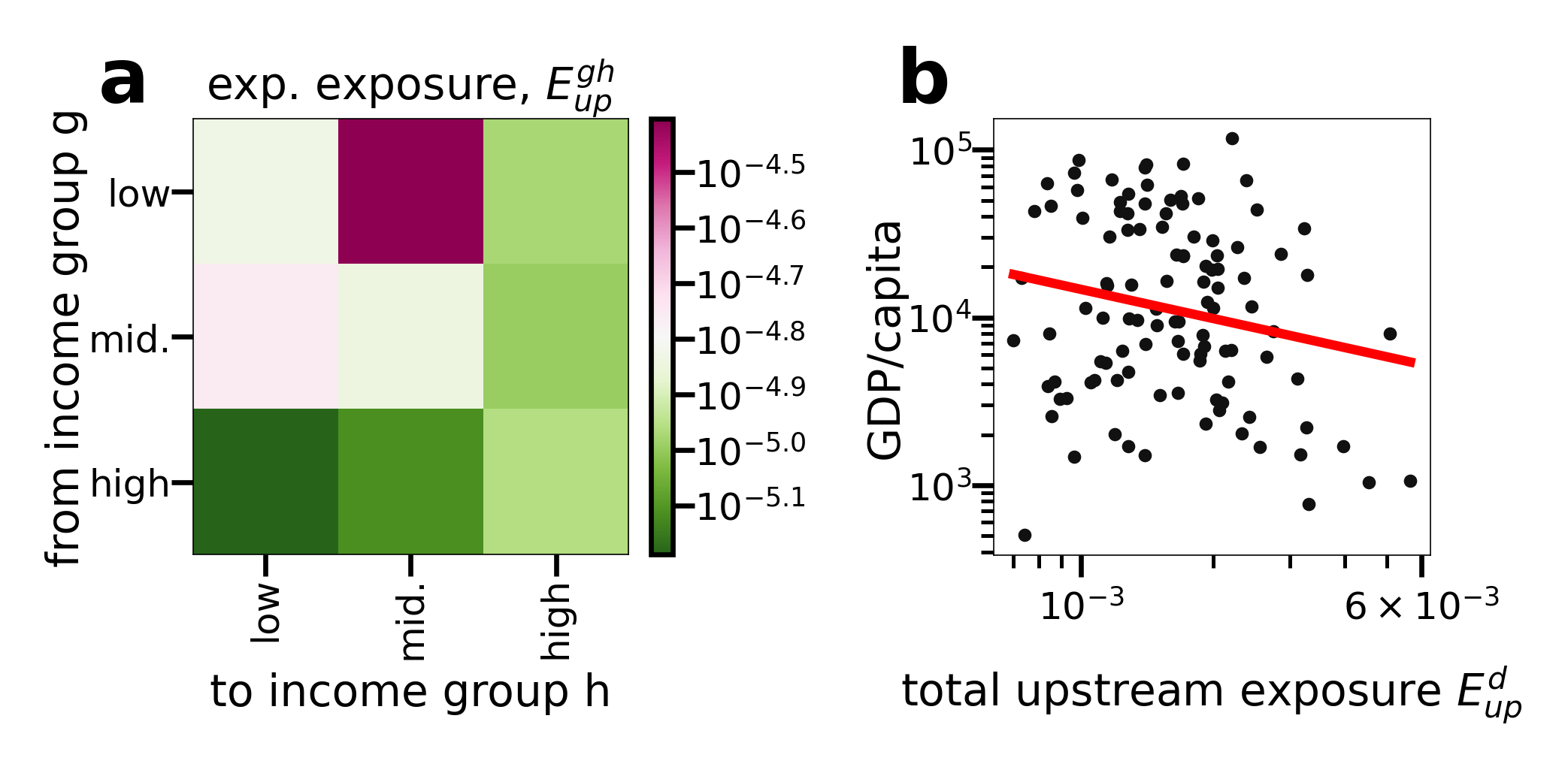}
		\caption{Country income and upstream exposure. 
			(a) Income group assortativity of expected upstream exposure. Distress ($\log$-scale) between firms separated into low-, middle- and high-income groups based on their country's GDP per capita. 
			Large exposures are mostly between low and middle income countries, high income countries do not experience much exposure. SI Table \ref{tab:SI:country_names} lists the income group for each country.
			(b) GDP per capita plotted against total upstream exposure $E^d_{up}$. A significant negative correlation of $r=-0.20, p <0.04$ highlights that higher exposure is connected to lower income per capita. The red line represents the log-log ordinary least squares regression fit to the data of form $y \sim x^{-0.58}$.}
		\label{fig:GDPassortativity_up}
	\end{figure}

	We show the average upstream distress exposure matrix between firms in low, middle and high income countries in SI Fig~\ref{fig:GDPassortativity_up}a.  It shows that firms in low and middle income countries affect each other more than any other pair of groups. High income countries experience little and create little exposure. 
	
	We plot the total upstream exposure $E^d_{up}$ with GDP per capita of the countries in SI Fig~\ref{fig:GDPassortativity_up}b. It shows that these two quantities are negatively correlated ($r=-0.16, p<0.04$), reflecting the fact that countries with lower GDP per capita receive more upstream imported distress, however with a weaker effect and lower significance.
	
\begin{figure*}[ht]
		\centering
		\includegraphics[width=0.33\textwidth]{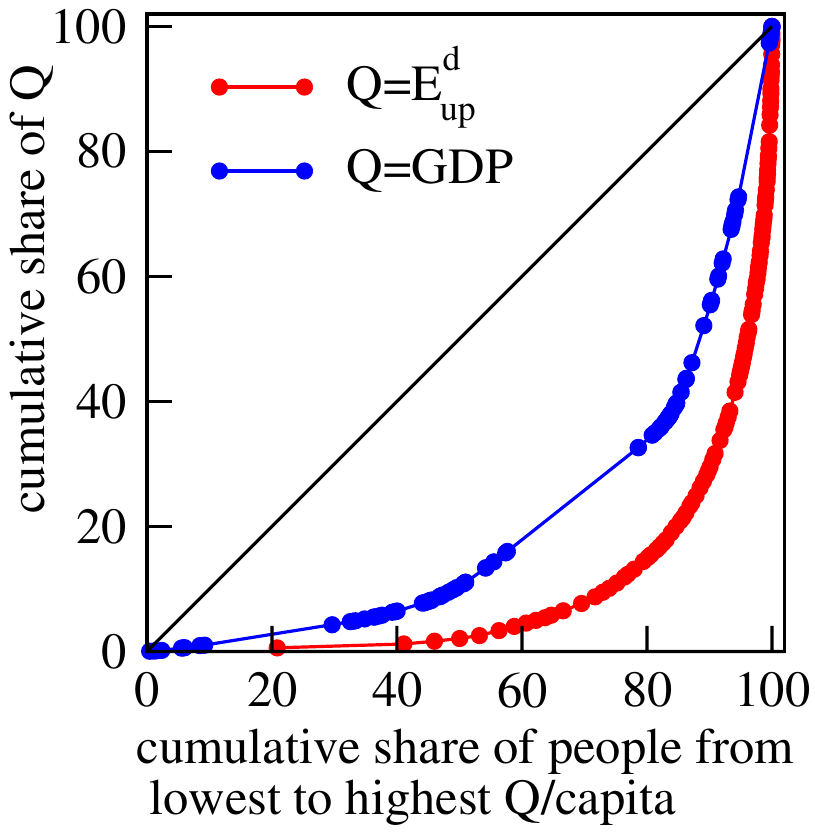}
		\caption{Lorenz curves for total upstream imported distress and GDP of all the countries. The red (blue) line shows the proportion of risk (GDP) that is assumed by the lowest exposed (poorest) x\% of people globally. For perfectly equally distributed exposure (wealth) the curve coincides with the diagonal, inequality is higher if the area between the curve and the diagonal increases. Exposure to economic shock is distributed more unequally than GDP. The area between the diagonal and the curve is proportional to the Gini index. We find Gini coefficients of 0.81 and 0.59 for $E^d_{up}$ and GDP, respectively.
		}
		\label{fig:inequality_up}
	\end{figure*}

	We investigate the concentration of exposure to upstream cascades with the Lorentz curve in SI Fig~\ref{fig:inequality_up}. Again, $E^d_{up}$ is concentrated more strongly than GDP with a Gini coefficient of $0.81$, compared to a Gini coefficient of $0.59$ for GDP.

	\subsection*{SI Text 8: Relationship between aggregate firms degree and macroeconomic variables}
\begin{figure}[ht]
		\centering
		\includegraphics[width=0.8\linewidth]{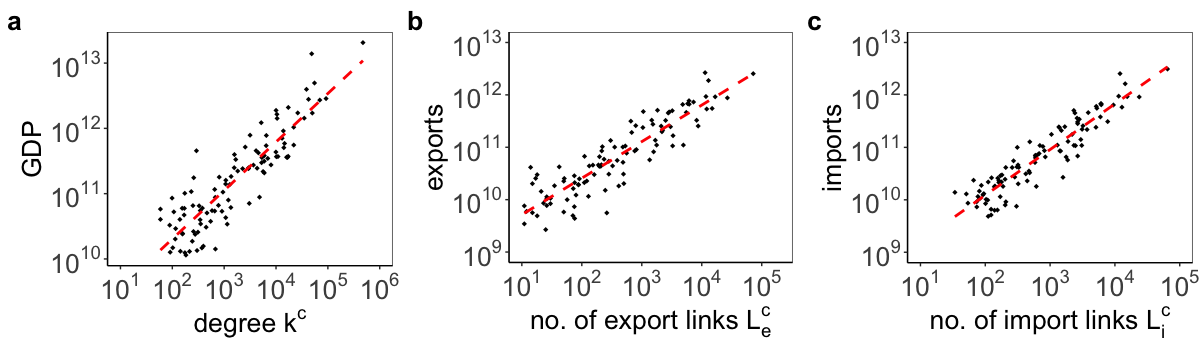}
		\caption{Comparison of country-aggregated degree and macroeconomic variables.
			(a) Total degree $k^c$ plotted against GDP. The red dotted line represents $GDP \sim ( k^c)^{0.74}$
			(b) Number of links out of a country $L_e^c$ compared to its export volume. The red dotted line represents Exports $\sim ( L_e^c)^{0.69}$
			(c) Number of links into a country $L_i^c$ compared to its import volume. The red dotted line represents Imports $\sim ( L_i^c)^{0.87}$.
			All in current USD. All investigated quantities correlate strongly and suggest a relationship between size and representation in the supply network.}
		\label{fig:weight}
	\end{figure}

	Our dataset lacks information on the traded volumes (i.e. link weights) and firm level information such as revenue. We assume equal weights on all links ($A_{ij}\in \{0,1\}$) and verify that this approximation is justified on the aggregate level by comparing it to macroeconomic variables. We calculate the total degree $k^c$, total export links $L_e^c$, and total import links  $L_i^c$ of firms in a country.
	The total degree  for a country $c$, is measured as the total number of links of all firms in country $c$. $L_e^c$ represents the total number of outgoing links of all firms in $c$ to firms in all other countries. Similarly, $L_i^c$ represents the total number of incoming links of all firms in the $c$ from firms in all other countries.  SI Figure~\ref{fig:weight} shows that these quantities are strongly correlated with GDP, total exports and total imports with  
	the Pearson correlation coefficients $r=0.87$, $p<10^{-15}$, $r=0.77$, $p<10^{-15}$ and  $r=0.86$, $p<10^{-15}$ respectively.
	
	\subsection*{SI Text 9: Results for states in the USA}
\begin{figure}[ht]
	    \centering
		\includegraphics[width=0.75\linewidth]{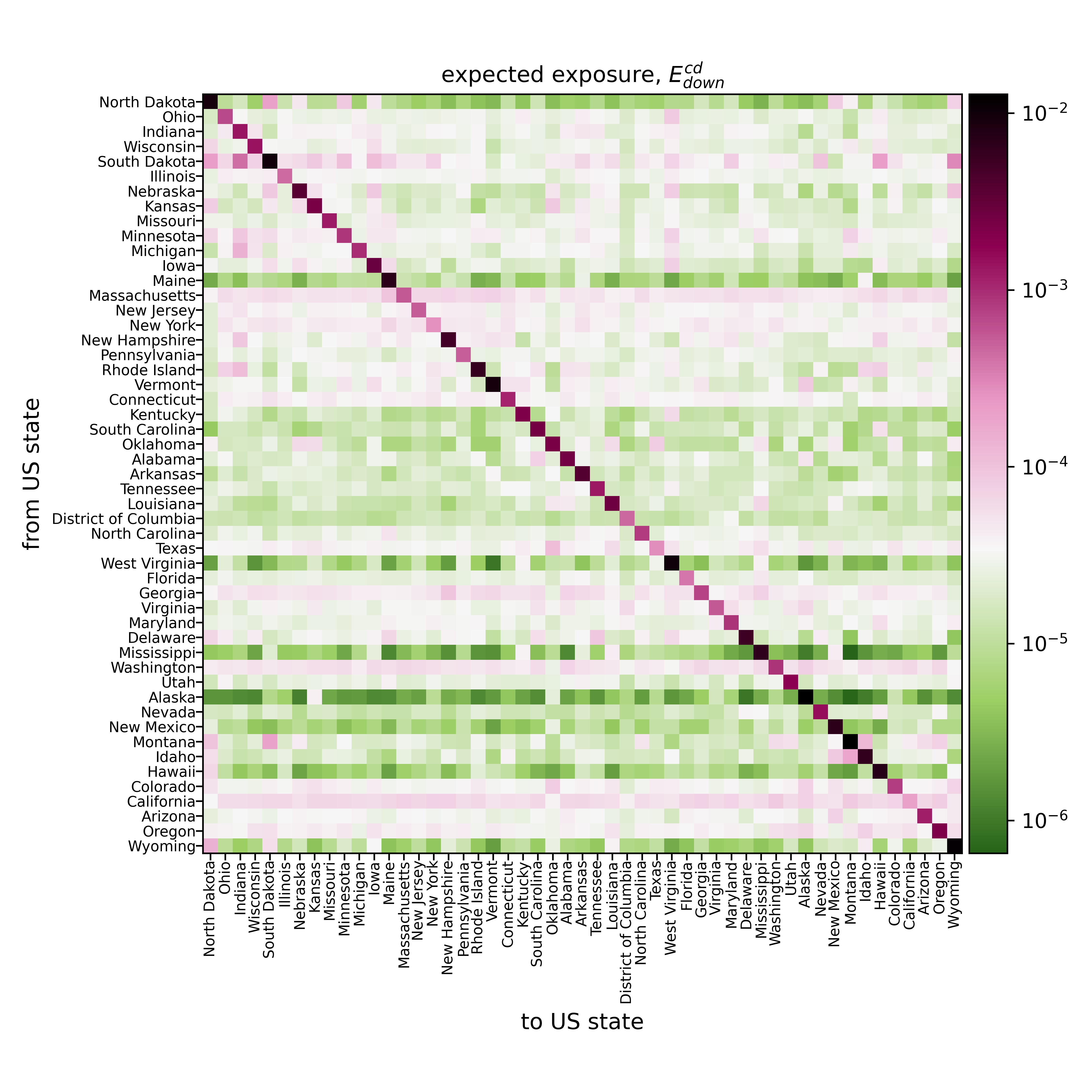}
		\caption{Exposures between US states in the national supply network, aggregated to US state level. 
			The expected fraction of the economy affected in US state $d$, subsequent to an arbitrary firm default in state $c$, $E^{cd}_{down}$, plotted on a logarithmic scale. The countries are sorted by US region (as defined by the US census bureau) in the following order: Midwest, Northeast, South, West. The values are highest along the diagonal, but in contrast to the to the global supply network the there are no regional blocks of high exposure. This can be explained by the history, as well as the physical and economic geography of the US.}
		\label{fig:heatmap_US}
	\end{figure}

	Here we study downstream shock propagation in the national supply network of the USA, aggregated to the 50 states of the USA. In SI Fig.~\ref{fig:heatmap_US} we plot the exposure matrix $E^{cd}_{down}$ between US states, sorted according to the regional classification by the US census bureau. Here, similarly to the exposures in the global supply network, the strongest exposures are along the diagonal --within the respective states--. However, we do not observe a pronounced block structure, indicating that exposures are not stronger within geographic regions. This can be explained by less pronounced regional differences, a shared cultural history and a lack of trade barriers within the US. The more pronounced structure is that large, rich states create a lot of exposure to the rest of the country, visible as bright and magenta horizontal lines.
	
\begin{figure}[htbp]
		\centering
		\includegraphics[width=0.5\linewidth]{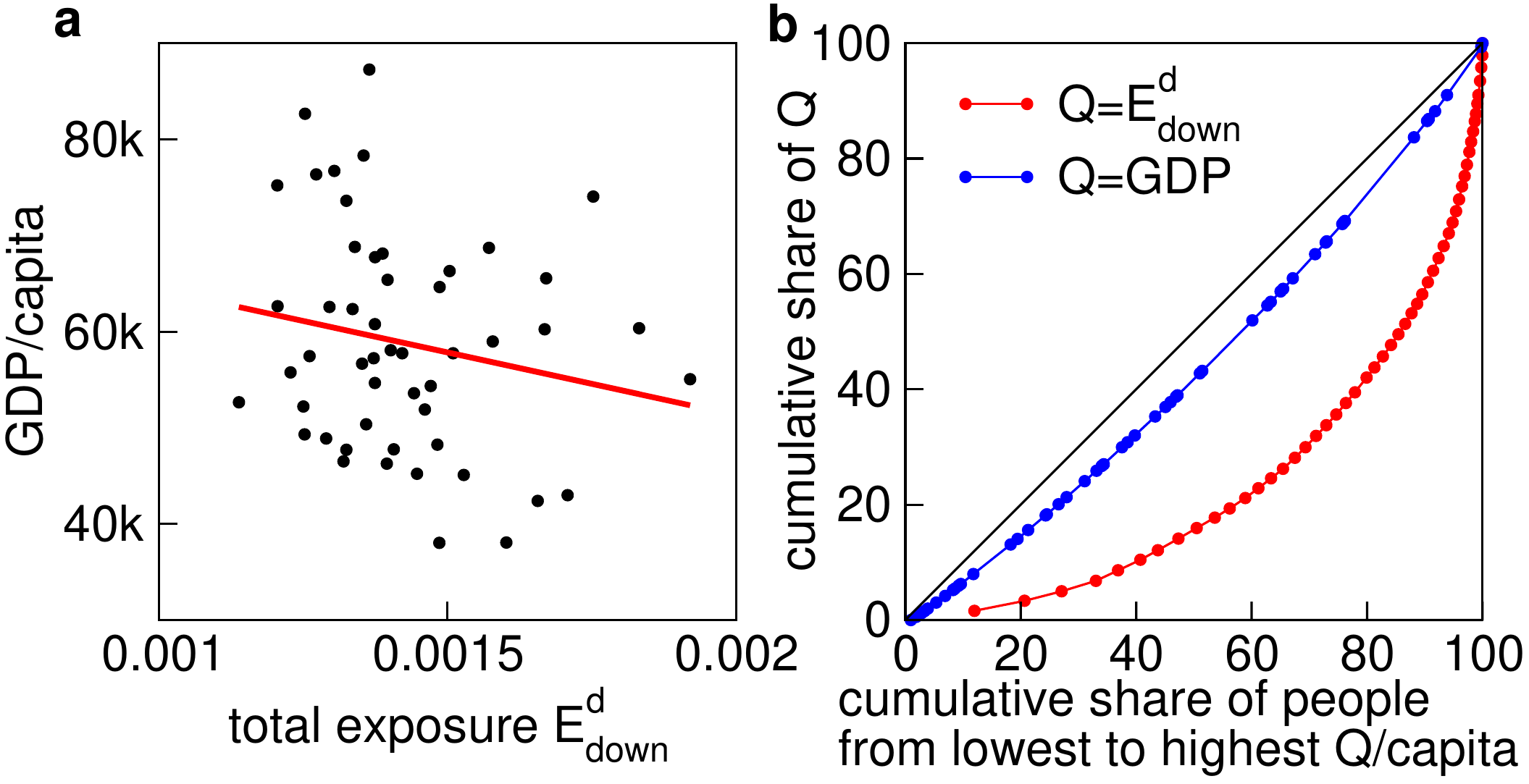}
		\caption{(a) Variation of total exposure with state GDP per capita of USA. The red line represents the  best  linear fit to  the  data  of  form $y = 0.08 - 13.1 x$ . (b) Lorenz curves for total exposure and GDP of all the states of USA. The Gini coefficients $=0.54$ and $0.11$ for total exposure $E^d_{down}$ and GDP respectively. GDP data are obtained from Bureau of Economic Analysis \url{https://www.bea.gov}, and population data are obtained from U.S. Census Bureau, Population Division \url{https://www.census.gov}.}
		\label{fig:inequality_usa}
	\end{figure}

	SI Figure~\ref{fig:inequality_usa}~(a) shows that the states with lower GDP per capita receive more total exposure $E^d_{down}$ than states with higher GDP per capita. Albeit the correlation being insignificant ($r=-0.16, p=0.28$), the data shows a weak negative relationship between GDP per capita and $E^{d}$, as indicated by the red trend line in SI Fig.~\ref{fig:inequality_usa}~(a).
	We investigate the inequality in total exposure $E^d_{down}$ and GDP for US states with the Lorentz curves in SI Fig.~\ref{fig:inequality_usa}~(b). The Gini coefficients are found to be $=0.54$ and $ 0.11$ for total exposure and GDP, respectively. This indicates that, albeit less than for the global supply network, total downstream exposure is more concentrated than GDP among states in USA. We conclude that, despite lower inter-state heterogeneity we find visually similar results to the global supply network.
	
\end{document}